\documentstyle[preprint,eqsecnum,aps]{revtex}
\begin{document}
\draft
\preprint{IASSNS-HEP-94/90}
\title{Fractional Statistics in One-Dimension: View From An Exactly
Solvable Model}
\author{Z. N. C. Ha}
\address{School of Natural Sciences, Institute for Advanced Study, \\
Princeton, New Jersey 08540}
\date{October 25, 1994}
\maketitle
\begin{abstract}
One-dimensional fractional statistics is studied using the
Calogero-Sutherland model (CSM) which describes a system of non-relativistic
quantum particles interacting with inverse-square two-body potential on a
ring.  The inverse-square exchange can be regarded as a pure statistical
interaction and this system can be mapped to an ideal gas
obeying the fractional exclusion and exchange statistics.
The details of the exact calculations of the dynamical correlation
functions for this ideal system is presented in this paper.
An effective low-energy one-dimensional ``anyon''
model is constructed; and its correlation functions
are found to be in agreement with those in the CSM; and this
agreement provides an evidence for the equivalence of the first- and
the second-quantized construction of the 1D anyon model at least in the
long wave-length limit. Furthermore, the finite-size scaling applicable to
the conformally invariant systems is used to obtain the complete set of
correlation exponents for the CSM.
\end{abstract}
\pacs{PACS 05.30-d, 71.10.+x, 71.27.+a, 73.40.H}
\narrowtext

\section{INTRODUCTION}
\label{sec:intro}

The fractional statistics in low-dimensional $(< 3)$ quantum systems
had been a subject appreciated only by a few \cite{begin,lm}; however,
the discovery of the quantum Hall effect \cite{fqhexp} has perhaps
changed the status of the subject forever, promoting and reshaping
it into one of the most profound as well as popular field in modern
physics and into a necessary conceptual tool in condensed
matter physics.

The fractional statistics is usually discussed in the context of
two-dimensional systems where the adiabatic transport of a test particle
around the others can be used to determine the statistics
independent of the dynamical nature of the interacting system.
In one-dimension (1D), however, the dynamics and the kinematics
can not be decoupled in an unambiguous way (i.e., an exchange
necessarily involves a scattering), and the assignment of the
statistics to 1D particles is largely a matter of taste.
On the other hand, a good taste will yield fruitful concepts and tools.

In this paper I study the 1D fractional statistics using a specific
model called Calogero-Sutherland Model (CSM) \cite{csm} which
describes a systems of non-relativistic quantum
particles interacting with a pairwise potential that falls off as
inverse-square of the distance between the particles.
One of the most important practical features of this model is
that the inverse-square potential can be regarded
as a pure statistical interaction and the model maps to an
ideal gas of particles obeying the fractional statistics
\cite{duncan94,ha94}.  Usually, the first step in understanding a
general class of interacting Fermi system known as Fermi
liquid is to study the ideal Fermi gas which gives rise to
important concepts and tools such as the Fermi surface.
Much in the same spirit I start with
the simplest 1D system (i.e., the CSM) obeying the fraction statistics.

First, I list here some of the recent developments.
A lattice cousin of the CSM called Haldane-Shastry Model (HSM),
which corresponds to the $SU(2)$ Heisenberg spin chain with
the inverse-square instead of the usual nearest neighbor exchange,
has triggered a surge of interest in this class of models \cite{hsm}.
The HSM is known to possess the quantum symmetry algebra known
as Yangian \cite{yangian} and can be considered as a model of
ideal $SU(2)$ spinon gas obeying the semionic
fractional statistics \cite{duncan92}.
The $SU(n)$ versions of the CSM \cite{hahal92} and
the HSM \cite{hahal92,kawa} now exist and
in particular the spectrum of $SU(n)$ HSM is known to possess
the Bethe ansatz string structure \cite{string}.
The list goes on, but I concentrate on the $U(1)$ CSM in this paper.

The CSM is intimately related to the circular ensembles in random matrix
theory first introduced by Freeman Dyson \cite{dyson}.
In particular, the eigenvalue distribution functions for the
orthogonal, unitary, and symplectic random matrices correspond to
the ground state wavefunctions of the CSM at the interaction
parameters $\lambda = 1/2$, $1$ and $2$, respectively.
($\lambda = 1$ case corresponds to the free Fermi gas.) Some static
correlation functions of the CSM can be calculated using the techniques
developed for the random matrices \cite{mehta}.

More recently, Simons {\it et. al.} have been successful in mapping the CSM
to the matrix model and to the non-linear sigma model where the
supersymmetric algebra is applicable \cite{efetov}, and
therefore are able to calculate the
dynamical density-density correlation functions (DDDCF) for the CSM
at $\lambda = 1/2$, $1$ and $2$ \cite{simon}.
Haldane and Zirnbauer using the similar method
calculate the one-particle Green's function at $\lambda = 2$
(i.e., the symplectic case) \cite{halz94}.
Forrester \cite{forrester} also calculates some static correlation
functions at integer interaction parameters using a generalized form
of the celebrated Selberg integral formula \cite{form}.

It turns out that the eigenstates of the CSM can be written in
terms of Jack polynomials \cite{duncan94,forr3}
whose known algebraic properties provide
a powerful and direct method for calculating the most general
correlation functions.
Recently, the author has been successful in calculating the exact
DDDCF and one-particle Green's
function at arbitrary rational interaction parameters \cite{ha94}.
The method employed is new and is one of the main subjects in this paper.

The exact calculations of the correlation functions further provide
conclusive evidences of the inherent fractional exclusion
and exchange statistics embodied in the CSM \cite{ha94}.
I also make a direct connection of this model to the
edge states of the fractional quantum Hall droplet by first constructing
an effective low-energy one-dimensional ``anyon'' fluid model based
on a general gauge invariance argument, and then showing that the dynamical
correlation functions agree with those of the CSM.  This connection
between the CSM and the edge states
has previously been suggested by various people \cite{csmedge}.

This paper is organized in the following way.  In Section
\ref{sec:eigenstates} I show how to construct general
eigenstates of the CSM, and in Section \ref{sec:Jack} some properties of
the Jack polynomials are introduced. In Section \ref{sec:fractional}
some key aspects of the exclusion and the exchange statistics
are discussed. I show how to calculate the DDDCF
and the one-particle Green's function
in Section \ref{sec:ddcf} and \ref{sec:gf}.
In Section \ref{sec:edge} the ``harmonic-fluid''
description of the anyon fluid is constructed and the dynamical
correlation functions calculated; and I further
show that they agree with those
of the CSM.  This agreement provides an explicit connection
between the CSM and the system of coupled left- and right-edges
of the fractional quantum Hall effect and further shows the equivalence
between the first- and the second-quantized construction of one-dimensional
``anyon'' gas at least in the long-wavelength limit.  In
Section \ref{sec:finite} the finite-size scaling as applied to the
conformally invariant systems is used to obtain the complete set of
correlation exponents. I also discuss some aspect of the lattice cousins of
the CSM in Section \ref{sec:lattice}.

\section{Eigenstates of Calogero-Sutherland Model}
\label{sec:eigenstates}

In this section I introduce the CSM and show how to construct
the general eigenstates of the model following Sutherland \cite{suther}.
First, the Hamiltonian for the CSM on a ring of length $L$ is given by
\begin{equation}
H = -\sum_{j=1}^N {\partial^2 \over \partial x_j^2} +
\sum_{j<l}{2\lambda(\lambda - 1) \over d^2(x_j - x_l)},
\label{hamil}
\end{equation}
where $\hbar^2/2m = 1$ and $d(x) = (L/2\pi) |\sin(\pi x/L)|$.
The ground state wavefunctions of the model at integer values of
$\lambda$ corresponds to 1D versions
of Laughlin's wavefunctions \cite{lwave} and are given by
\begin{equation}
\Psi_0 = \prod_{i<j} (z_i - z_j)^\lambda \prod_k z_k^{J_0},
\end{equation}
where the current $J_0 = -\lambda(N-1)/2$ and $z_j = \exp(i2\pi x_j/L)$.
When $0 < \lambda < 1$ there is another possible ground state with
power $1-\lambda$; however, only the solution with power $\lambda$
will be considered for reasons of continuity with $\lambda > 1$ solutions.

The excited states of this model are constructed by multiplying some
symmetric polynomials to the ground state wavefunction, and this
construction is analogous to that of the gapless edge excitations of the
quantum Hall effect \cite{edge}.
A general excited state $\Psi_{\bf n}^\lambda =
\Psi_0 J_{\bf n}^\lambda$ is labeled
by the quantum numbers
${\bf n} = (n_1, n_2, \ldots, n_N)$,
and $J^\lambda_{\bf n}$ satisfies the following
new eigenvalue equation
\begin{equation}
\tilde H J_{\bf n}^\lambda =
E_{\bf n} J_{\bf n}^\lambda,
\label{neweigen}
\end{equation}
where $\tilde H = H_0 + \lambda H_1$, and
\begin{eqnarray}
H_0 & = & \sum_{j=1}^N (z_j\partial_{z_j})^2, \\
H_1 & = & \sum_{j<k}
\frac{z_j + z_k}{z_j - z_k} (z_j\partial_{z_j} - z_k\partial_{z_k}).
\end{eqnarray}
The eigenstates of the new Hamiltonian $\tilde H$ are represented
in terms of the following bosonic basis states
\begin{equation}
\Phi({\bf n}) = \sum_P \prod_{j=1}^N z_j^{n_{P_j}},
\end{equation}
where the sum extends over all permutations of the integer
set ${\bf n}$ which can be considered as a set
of bosonic quantum numbers with no restrictions on their values.
Since $\Phi({\bf n})$ does not depend
on the ordering of the quantum numbers, let $n_1 \ge n_2 \ge \ldots \ge n_N$
without loss of generality.
These symmetric polynomials form a complete basis.

The action of $\tilde H$ on $\Phi({\bf n})$ can be easily calculated
and are given by
\begin{eqnarray}
H_0 \Phi({\bf n}) & = & \left( \sum_{j=1}^N n_j^2 \right) \Phi({\bf n}), \\
H_1 \Phi({\bf n}) & = & \sum_{j < k} (n_j - n_k) \left(
\Phi({\bf n}) + 2 \sum_{s=1}^{n_j-n_k-1} \Phi(\ldots,n_j-s,\ldots,n_k+s,
\ldots)\right).
\label{h1}
\end{eqnarray}
$H_0$ generates only the basis state $\Phi({\bf n})$ itself while $H_1$ is
responsible for generating a family of states
$\Phi(\ldots,n_j-s,\ldots,n_k+s,\ldots)$ which are
obtained from $\Phi({\bf n})$ by all possible pairwise ``squeezing'' of
the quantum numbers.  If a state is generated from another by
squeezing a pair of quantum numbers by one unit
(i.e., $n_j \rightarrow n_j - 1$, $n_k \rightarrow n_k +1 $ for
$n_j - n_k \ge 2$), then I call the former
``daughter-state'' and the latter its ``mother-state.''

The family of states can be organized into {\it levels} such
that the members of a given level are mutually not related
or {\it unreachable} (i.e.,
no mother-daughter relationship exists between the members in the same
level)
and the daughters of a member from a given level always
belong to a {\it lower} level in the family.
The highest-level mother-state denoted by $|\nu\rangle_1$, where
$\nu$ is a level index and is equal to the total number of levels
in the family, generates the entire family of daughter-states
which are denoted by $|\mu\rangle_\alpha$
where $1 \le \mu < \nu$ and $\alpha$ an index for the states in
the $\mu$th level.
While the order of the levels in the family can be uniquely determined,
the order within a level is quite arbitrary.

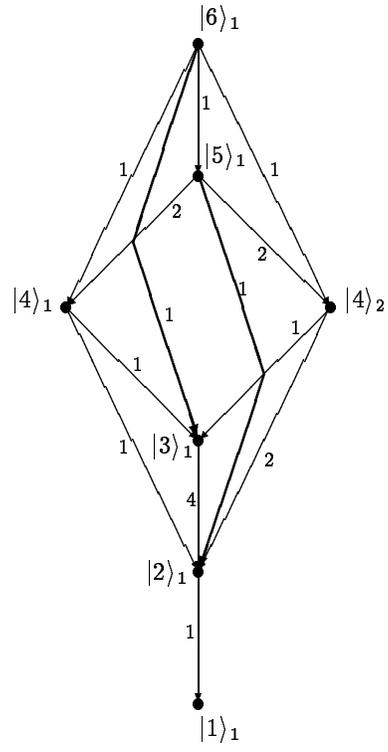
\begin{figure}
\caption{A level diagram for a family with the highest-level mother-state
given by $|6\rangle_1 = \Phi(4,3,1,0)$. The arrows connect the
mother states to their daughter states that are generated by
squeezing on the pairs of quantum numbers by one unit.}
\label{tree}
\end{figure}

To provide an illustration of the above mentioned family structure,
I give the following example. Let the highest-level mother-state be
$|6\rangle_1 = \Phi(4,3,1,0)$. Then, the members of the family are
$|5\rangle_1 = \Phi(4,2,2,0)$, $|4\rangle_1 = \Phi(4,2,1,1)$,
$|4\rangle_2 = \Phi(3,3,2,0)$, $|3\rangle_1 = \Phi(3,3,1,1)$,
$|2\rangle_1 = \Phi(3,2,2,1)$, and $|1\rangle_1 = \Phi(2,2,2,2)$.
A pictorial representation of the family structure is shown in
Fig. \ref{tree} and I call it a {level diagram}.
In Fig. \ref{tree} the states are represented by dots and
each arrow connects a state $A$ to $B$ where $B$ is {\it reachable} from $A$
by {\it squeezing} on a pair of quantum numbers for $A$ by one unit
(i.e., connects a mother and her daughter).  The levels are ordered from top
to bottom, from the most squeezable to unsqueezable states,
such that the highest-level mother-state is at the top; and
therefore the arrows always point downwards and never upwards.
A set of arrows that is topologically equivalent to a directed line forms
a {\it path}.
A {\it weight} $\cal W$ of an arrow is defined to be $m(n_i)m(n_j)$, where
$n_i$ and $n_j$ are the two quantum numbers squeezed to produce a
daughter and $m(l)$ the multiplicity of $l$ in the quantum number set
specifying the mother-state.  In Fig. \ref{tree} the numbers next
to the arrows are the corresponding weights.
The states in the same level are
not connected; hence, the Hamiltonian is diagonal in that subspace.

A {\it subfamily} of a family can also be constructed by starting from a given
state, which would be the highest-level mother-state of that subfamily,
and grouping all her {\it reachable} off-springs.
The total number of {\it subfamilies} is therefore equal to the {\it dimension}
(i.e., the total number of basis states) of the family.

The lowest-level states (the unsqueezable states) clearly have to be of
one of the two types: (I) $\Phi(\ldots,m,m,m,m,m,\ldots)$ or
(II) $\Phi(\ldots,m,m,m,m,m-1,m-1,m-1,m-1,\ldots)$.  The type-I state
corresponds to the ground state up to a global Galilean boost and
the type-II to states with a single hole excitation. These one-hole states
are eigenstates of $\tilde H$ and, furthermore, since the following
superposed state
\begin{equation}
\Psi(x) = \prod_{j=1}^N (z - z_j),
\label{onehole}
\end{equation}
where $z = \exp(i 2\pi x/L)$,
can be expanded purely in terms of the one-hole eigenstates, $\Psi(x)$
describes a state with a hole localized at $x$.

The matrix representation of $\tilde H$ acting on the partially ordered state
space is always triangular since the action of $\tilde H$ on a given
state always generates states belonging to lower levels.
The eigenvalues, therefore, are simply given by
the diagonal matrix elements.  In particular the energy of an
eigenstate spanned by a family with the highest-level mother-state
$\Phi(\bf n)$ is given by
\begin{equation}
E_{\bf n}^{(0)} = \sum_{j=1}^N n_j^2 + \lambda \sum_{j<k} (n_j - n_k).
\label{eng}
\end{equation}
The off-diagonal elements are given by
\begin{equation}
\mbox{}_\alpha\langle \mu|\tilde H|\nu\rangle_\beta
= \left\{ \begin{array}{ll}
          \left[\sum_{\cal P} \left(\prod_{i\in {\cal P}}
          {\cal W}_i\right) \right]
          E_{\bf n}^{(1)} & \mbox{if $\nu > \mu$}, \\
          0 & \mbox{if $\nu \le \mu$},
         \end{array}
  \right.
\end{equation}
where $E_{\bf n}^{(1)} = 2\lambda \sum_{j<k} (n_j - n_k)$, and
the sum is over all possible path ${\cal P}$ from $|\nu\rangle_\beta$
to $|\mu \rangle_\alpha$ and the product over all the weights ${\cal W}_i$
of the intermediate arrows belonging to $\cal P$.

For the example given in Fig. \ref{tree},
$\tilde H$ is represented by
\begin{equation}
\tilde H = \left( \begin{array}{ccccccc}
\varepsilon^{(1,1)}_{(1,1)}&\varepsilon^{(1,1)}_{(2,1)}&
\varepsilon^{(1,1)}_{(3,1)}&\varepsilon^{(1,1)}_{(4,1)}&
\varepsilon^{(1,1)}_{(4,2)}&\varepsilon^{(1,1)}_{(5,1)}&
\varepsilon^{(1,1)}_{(6,1)} \\
0&\varepsilon^{(2,1)}_{(2,1)}&
\varepsilon^{(2,1)}_{(3,1)}&\varepsilon^{(2,1)}_{(4,1)}&
\varepsilon^{(2,1)}_{(4,2)}&\varepsilon^{(2,1)}_{(5,1)}&
\varepsilon^{(2,1)}_{(6,1)} \\
0&0&\varepsilon^{(3,1)}_{(3,1)}&\varepsilon^{(3,1)}_{(4,1)}&
\varepsilon^{(3,1)}_{(4,2)}&
\varepsilon^{(3,1)}_{(5,1)}&\varepsilon^{(3,1)}_{(6,1)} \\
0&0&0&\varepsilon^{(4,1)}_{(4,1)}&0&
\varepsilon^{(4,1)}_{(5,1)}&\varepsilon^{(4,1)}_{(6,1)} \\
0&0&0&0&\varepsilon^{(4,2)}_{(4,2)}&\varepsilon^{(4,2)}_{(5,1)}&
\varepsilon^{(4,2)}_{(6,1)} \\
0&0&0&0&0&\varepsilon^{(5,1)}_{(5,1)}&\varepsilon^{(5,1)}_{(6,1)} \\
0&0&0&0&0&0&\varepsilon^{(6,1)}_{(6,1)}
\end{array} \right),
\end{equation}
where $\varepsilon^{(\mu,\alpha)}_{(\nu,\beta)} =
\mbox{}_\alpha\langle \mu|\tilde H|\nu\rangle_\beta$.
For example, $\varepsilon^{(3,1)}_{(6,1)} = 7 \times E^{(1)}_{(3,3,1,1)}$
since there are five different ways to get from $|6\rangle_1$ to
$|3\rangle_1$ with the following corresponding weights (see Fig. \ref{tree}):
(a) $|6\rangle_1 \stackrel{1} \rightarrow |4\rangle_1
\stackrel{1} \rightarrow |3\rangle_1$;
(b) $|6\rangle_1 \stackrel{1} \rightarrow |5\rangle_1
\stackrel{2} \rightarrow  |4\rangle_1
\stackrel{1} \rightarrow |3\rangle_1$;
(c) $|6\rangle_1 \stackrel{1} \rightarrow |3\rangle_1$;
(d)$|6\rangle_1 \stackrel{1} \rightarrow |4\rangle_2
\stackrel{1}\rightarrow |3\rangle_1$;
(e) $|6\rangle_1 \stackrel{1} \rightarrow |5\rangle_1
\stackrel{2} \rightarrow  |4\rangle_2 \stackrel{1} \rightarrow |3\rangle_1$.

The eigenenergy given by Eq. (\ref{eng}) plus the ground state energy
can be rewritten in terms of newly defined pseudomomenta $k_j$ as
\begin{equation}
E = {\hbar^2 \over 2m}\sum_{j=1}^N k_j^2,
\end{equation}
where
\begin{equation}
Lk_j = 2\pi I_j + \pi(\lambda-1)\sum_{l=1}^N \mbox{sgn}(k_j-k_l).
\label{ks}
\end{equation}
The quantum numbers $I_j$ are now distinct (half-odd) integers
and are related to $n_j$'s by $I_j = n_j + (N+1-2j)/2$.

The distribution of $k_j$ determined by Eq.\ (\ref{ks}) is used
to construct pictorial representations of the eigenstates called
motifs which are crucial for exposing the fractional statistics
obeyed by the elementary excitations of the model.  Detailed
discussion of this subject is given in Section \ref{sec:fractional}.

\section{Jack Symmetric Polynomials}
\label{sec:Jack}

The polynomial solutions of Eq. (\ref{neweigen}) in Section
\ref{sec:eigenstates} is
also known in mathematical literature as Jack polynomials
\cite{jack}. In fact,
Stanley \cite{stanley} has shown that the complete set of linearly
independent solutions of Eq. (\ref{neweigen}) is indeed
given by Jack polynomials up to global Galilean boosts (i.e.,
up to the factor $\prod_{j=1}^N z_j^J$ where $J$ is the current
and takes an arbitrary real number) \cite{ha94}.

For better readability this section is divided into two subsections:
the first introduces the conventional notations used in mathematical
literatures and the second some of the general properties of Jack polynomials.

\subsection{Introduction to notations}
{\it Partitions} are defined as sequences of non-negative integers
in non-increasing order and are used to label the symmetric polynomials.
They are denoted by bold-face Greek letters as
\begin{equation}
{\mbox{\boldmath $\kappa$}} = (\kappa_1, \kappa_2, \ldots, \kappa_N),
\end{equation}
where $\kappa_1 \ge \kappa_2 \ge \ldots \ge \kappa_N$.
Non-zero $\kappa_j$ are
called {\it parts} of {\boldmath $\kappa$}
whose {\it length} (i.e., the total number
of non-zero {\it parts}) is denoted by $\ell({\mbox{\boldmath $\kappa$}})$.
The {\it weight} of the partition
is defined by $|{\mbox{\boldmath $\kappa$}}| =
\sum_{j=1}^{\ell({\mbox{\boldmath $\kappa$}})} \kappa_j$.
If $\kappa_1+\ldots+\kappa_i \ge \mu_1+\ldots+\mu_i$ for all $i \ge 1$,
then ${\mbox{\boldmath $\kappa$}} \ge {\mbox{\boldmath $\mu$}}$.

Young diagram ${\cal D}({\mbox{\boldmath $\kappa$}})$ is used
to graphically represent a partition:
${\cal D}({\mbox{\boldmath $\kappa$}}) = \{ (i,j): 1\le i \le
\ell({\mbox{\boldmath $\kappa$}}),\ 1\le j \le \kappa_i \}$.
The cell labeled by
$(i,j)$ is situated in the $i$-th row and the $j$-th column of the Young
diagram. The diagram of {\boldmath $\kappa$}, therefore,
consists of $\ell({\mbox{\boldmath $\kappa$}})$
rows of lengths $\kappa_j$.

A {\it conjugate} of {\boldmath $\kappa$} is denoted by
${\mbox{\boldmath $\kappa$}}' =
(\kappa_1', \kappa_2', \ldots)$  and corresponds to a partition
whose diagram is obtained by changing all the rows of
${\cal D}({\mbox{\boldmath $\kappa$}})$ to columns in
non-increasing order from the left to right.
For example, the conjugate of ${\mbox{\boldmath $\kappa$}} = (5,2,2,1)$
is ${\mbox{\boldmath $\kappa$}}' = (4,3,1,1,1)$.
Now, the following simple but useful identity can be derived \cite{macbook1}
\begin{equation}
n({\mbox{\boldmath $\kappa$}}) \equiv
\sum_{i=1}^{\ell({\mbox{\boldmath $\kappa$}})} (i-1)\kappa_i =
\sum_{i=1}^{\ell({\mbox{\boldmath $\kappa$}}')}
{\kappa_i'\choose 2}.
\label{simpleid}
\end{equation}
In order prove Eq. (\ref{simpleid}) every cell in the $i$th row of
${\cal D}({\mbox{\boldmath $\kappa$}})$ is filled in
with an integer $i-1$.
Since $n({\mbox{\boldmath $\kappa$}})$ corresponds to the sum
of all the integers
in the diagram, the two different expressions for
$n(\mbox{{\boldmath $\kappa$}})$
are obtained depending on whether the numbers in each row
or column are summed first.

For a given cell $s=(i,j)$ of a diagram
${\cal D}({\mbox{\boldmath $\kappa$}})$
there are corresponding {\it arm-length} $a(s)=\kappa_i - j$,
{\it arm-colength} $a'(s) = j - 1$, {\it leg-length}
$l(s) = \kappa_j' - i$, and {\it leg-colength} $l'(s) = i - 1$.
The {\it upper} and {\it lower hook-lengths} are defined,
respectively, as
\begin{eqnarray}
h^*_{\mbox{\boldmath $\kappa$}}(s)& = &l(s) + {1 + a(s)\over \lambda}, \\
h_*^{\mbox{\boldmath $\kappa$}}(s)& = &l(s) + 1 + {a(s)\over \lambda}.
\end{eqnarray}

\subsection{General properties of Jack polynomials}
The symmetric polynomials are indexed by the partitions. The bosonic
basis functions $\Phi({\bf n})$ of the CSM are called the
monomial symmetric functions and the quantum numbers $\bf n$ correspond
to the partitions defined in the previous subsection.  Since the quantum
numbers are allowed to be negative integers, the correspondence is only
up to some trivial constant translation or the global Galilean boost.
This restriction to non-negative integer parts is more of a convenience
than a restriction since the CSM Hamiltonian is invariant under
the global Galilean transformation.

I shall denote Jack symmetric polynomials as
$J^{1/\lambda}_{\bf\kappa} (z_1,\ldots,z_N)$ which are
the solutions of Eq. (\ref{neweigen}). If $\lambda = 1$, Jack
polynomials reduce to Shur functions which describe the excited states
of the free fermions.  At $\lambda = 0$, it becomes the monomial symmetric
function which is just the free bosonic wavefunction.
If $\lambda = 2$ or $1/2$, they are called the
zonal spherical functions. As $\lambda \rightarrow \infty$,
$J^{1/\lambda}_{\bf\kappa}$
reduce to the elementary symmetric functions.

One way of defining Jack polynomials is through the differential equation
(\ref{neweigen}).  The other is
based on the properties of the power-sum symmetric function
$p_{\mbox{\boldmath $\kappa$}} =
p_{\kappa_1}p_{\kappa_2}p_{\kappa_3}\ldots$, where
$p_{\kappa_\nu} = \sum_j z_j^{\kappa_\nu}$.  Define a bilinear scalar
product on the vector space of all symmetric functions of finite
degree as
\begin{equation}
\langle p_{\mbox{\boldmath $\kappa$}}, p_{\mbox{\boldmath $\mu$}}
\rangle_{1/\lambda} =
\delta_{{\mbox{\boldmath $\kappa$}},{\mbox{\boldmath $\mu$}}}
z_{\mbox{\boldmath $\kappa$}} \lambda^{-\ell({\mbox{\boldmath $\kappa$}})},
\end{equation}
where $z_{\mbox{\boldmath $\kappa$}} = \prod_{i \ge 1} i^{m_i} m_i !$, and
$m_i = m_i({\mbox{\boldmath $\kappa$}})$ is the number of parts
of {\boldmath $\kappa$} equal to
$i$.  Using this definition,
Macdonald \cite{macbook2} proved that there are unique symmetric functions
satisfying the following three properties:
\begin{enumerate}
\item {\it Orthogonality}: $\langle J_{\mbox{\boldmath $\kappa$}},
J_{\mbox{\boldmath $\mu$}} \rangle_{1/\lambda}
= \delta_{{\mbox{\boldmath $\kappa$}},{\mbox{\boldmath $\mu$}}}
j_{\mbox{\boldmath $\kappa$}}^\lambda$, where
$j_{\mbox{\boldmath $\kappa$}}^\lambda$ is the normalization constant.
\item {\it Triangularity}: $J_{\mbox{\boldmath $\kappa$}} =
\sum_{\mbox{\boldmath $\mu$}} v_{\mbox{\boldmath $\kappa\mu$}}
\Phi({\mbox{\boldmath $\mu$}})$, where $ v_{\mbox{\boldmath $\kappa\mu$}}
= 0$ unless ${\mbox{\boldmath $\kappa$}} \le {\mbox{\boldmath $\mu$}}$.
\item {\it Normalization}: If $|{\mbox{\boldmath $\kappa$}}| = n$, then
$v_{\mbox{\boldmath $\kappa\mu$}} = n!$, where
${\mbox{\boldmath $\kappa$}} = (\underbrace{1,1,\ldots,1}_n)$.
\end{enumerate}
$J_{\mbox{\boldmath $\kappa$}}$ are, then, constructed by Gram-Schmidt
orthogonalization relative to the scalar product on the ring of
polynomials. Stanley \cite{stanley} proved that the normalization
constant is given by
\begin{equation}
j_{\mbox{\boldmath $\kappa$}}^\lambda =
\prod_{s\in {\mbox{\boldmath $\kappa$}}}
h^*_{\mbox{\boldmath $\kappa$}}(s) h_*^{\mbox{\boldmath $\kappa$}}(s).
\end{equation}

There is another scalar product on which Jack polynomials are orthogonal:
\begin{eqnarray}
\langle {\mbox{\boldmath $\kappa$}} | {\mbox{\boldmath $\mu$}}
\rangle_{1/\lambda}
 & \equiv & A_N^2 \left( \prod_{j=1}^N \int_0^L \!\! dx_j \right)
\overline{J_{\mbox{\boldmath $\kappa$}}^{1/\lambda}(z_1,z_2,\ldots,z_N)}
J_{\mbox{\boldmath $\mu$}}^{1/\lambda}(z_1,z_2,\ldots,z_N) \prod_{i<j}
|z_i-z_j|^{2\lambda} \nonumber \\
\mbox{} & = & A_N^2 j_{\mbox{\boldmath $\kappa$}}^\lambda
\prod_{s\in {\mbox{\boldmath $\kappa$}}}
{N + a'(s)/\lambda - l'(s) \over N+ (a'(s) + 1)/\lambda -
(l'(s) + 1)} \delta_{{\mbox{\boldmath $\kappa$}},
{\mbox{\boldmath $\mu$}}},
\label{orth}
\end{eqnarray}
where $z_j = \exp(i2\pi x_j/L)$ and
$A_N^2 = (1/L)^N \Gamma^N(1+\lambda)/\Gamma(1+\lambda N)$, and
the bar over the polynomial denotes the complex conjugation.
Note also that the multidimensional integral above
is equal to $L^N$ times the constant term in
\begin{equation}
J_{\bf\kappa}^{1/\lambda}(1/z_1,1/z_2,\ldots,1/z_N)
J_{\bf\mu}^{1/\lambda}(z_1,z_2,\ldots,z_N)
\prod_{i\ne j} (1-\frac{z_i}{z_j})^\lambda .
\label{orth1}
\end{equation}
Eq. (\ref{orth}) has been conjectured first by Macdonald \cite{mac1}
and then later proved by himself \cite{macbook2}
and also by Kadell \cite{kadell}.

Since Jack polynomials span the vector space of symmetric functions, they
can be used to expand any symmetric functions.
This property is particularly useful in calculating the correlation functions
of the CSM as will be shown later in this paper.
Here are some of them \cite{hanlon,yan}:
\begin{eqnarray}
\sum_{i=1}^N z_i^n & = & {n \over \lambda}
\sum_{|{\mbox{\boldmath $\kappa$}}| = n}
{[0']^\lambda_{\mbox{\boldmath $\kappa$}} \over
j^\lambda_{\mbox{\boldmath $\kappa$}}}
J^{1/\lambda}_{\mbox{\boldmath $\kappa$}}(z_1,\ldots,z_N),
\label{power} \\
\prod_{j=1}^N (1-z_j)^a & = & \sum_{\mbox{\boldmath $\kappa$}}
{\{-a\}^\lambda_{\mbox{\boldmath $\kappa$}} \over
\lambda^{|\mbox{\boldmath $\kappa$}|} j^\lambda_{\mbox{\boldmath $\kappa$}}}
J_{\mbox{\boldmath $\kappa$}}^{1/\lambda}(z_1,\ldots,z_N),
\label{destruct}
\end{eqnarray}
where $[a]^\lambda_{\mbox{\boldmath $\kappa$}} =
\prod_{(i,j) \in {\mbox{\boldmath $\kappa$}}} \{ a + (j-1)/\lambda -(i-1)\}$,
and $\{a\}^\lambda_{\mbox{\boldmath $\kappa$}} =
\prod_{(i,j) \in {\mbox{\boldmath $\kappa$}}}
\{ a - \lambda(i-1) + (j-1)\}$.
The sum in Eq. (\ref{destruct}) extends over all possible partitions while
in Eq. (\ref{power}) it is restricted to partitions with
{\it weight} $|{\mbox{\boldmath $\kappa$}}| = n$.
The prime in $[0']^\lambda_{\mbox{\boldmath $\kappa$}}$
denotes that the product does not include
the cell $(0,0)$; otherwise the total
product is trivially equal to zero.
$J_{\mbox{\boldmath $\kappa$}}$ also satisfies
\begin{equation}
J_{\mbox{\boldmath $\kappa$}}^{1/\lambda}(w z_1,w z_2,\ldots,w z_N) =
w^{|\mbox{\boldmath $\kappa$}|}J_{\mbox{\boldmath $\kappa$}}^{1/\lambda}
(z_1,z_2,\ldots,z_N),
\label{homo}
\end{equation}
since Jack polynomials are homogeneous functions of degree
${|\mbox{\boldmath $\kappa$}|}$.

\section{Fractional statistics}
\label{sec:fractional}

I divide this section into two subsections. In the first (second) subsection
the fractional {\it exchange} ({\it exclusion}) statistics is
discussed in the context of one-dimensional models.
The exchange statistics in two-dimension is directly relevant to the
fractional quantum Hall effect and various people have
made contributions to this fascinating field \cite{qhe}.
In one-dimension, however, the definition of fractional exchange statistics
is rather obscure and incomplete with a possible exception of the CSM.
One the other hand, the definition of fractional exclusion statistics
is spatial dimension independent and is based on the structure of
the Hilbert space rather than the configuration space.
While the fermions obey the well-known Pauli exclusion principle,
more exotic particles may obey a ``generalized exclusion
principle.'' \cite{duncan92}

\subsection{Exchange Statistics}
The first full mathematical treatment of the fractional statistics is
due mainly to Leinaas and Myrheim who used the multiply connected
topological structure of the configuration space of collections
of identical particles to show the possibility of exotic statistics
in spatial dimension less than three \cite{lm}.  While their idea has been
extensively applied to two-dimensional systems especially in the
context of the fractional quantum Hall effect,
very little attention has been paid to the one-dimensional (1D) systems.
Perhaps the main difficulty in 1D systems is that the physical
exchanges of the particles
necessarily involve scattering processes and that there is no known
unique way of un-tangling the kinematic aspect of the fractional
statistics from the dynamical processes.

For integrable 1D quantum systems, however, there is a general consistency
condition known as the Yang-Baxter equation (YBE) which essentially
puts strong restrictions on the scattering matrices. Therefore,
intuitively the braiding of the particle ``trajectories'' in
one-dimension when properly defined may be given by the YBE \cite{frohlich}.
In fact the YBE is known to be intimately related to the
the knot theory and the braid groups \cite{ybe}.
Hence, perhaps the following
quantum Yang-Baxter equation should be interpreted as the
one-dimensional generalized braiding relations,
\begin{equation}
S_{12}(v-u)S_{13}(v)S_{23}(u) =
S_{23}(u)S_{13}(v)S_{12}(v-u),
\label{ybe}
\end{equation}
where $S_{ij}(v)$ is the scattering matrix in the tensor product of linear
vector spaces, $V\otimes V\otimes V$ and acts non-trivially only in
the $i$th and $j$th space, e.g. $S_{12}(v) = R(v)\otimes I$
where $R(v)$ is a matrix defined in $V\otimes V$.
The parameters $v$ and $u$ are called spectral parameters and
are equal to the rapidity differences between two colliding particles.
One can rewrite Eq. (\ref{ybe}) in terms of $\tilde R(v) =
PR(v)$ where $P$ denotes the transposition, $P x\otimes y
= y \otimes x$.  Define matrices $\tilde R_i(v) =
I\otimes \cdots \otimes \tilde R(v) \otimes \cdots \otimes I$
on $V\otimes \cdots \otimes V$, where $\tilde R(v)$ acts on
$i$-th and $i+1$-th spaces.  The matrices $\tilde R_i(v)$
satisfy the following relations
\begin{eqnarray}
\tilde R_i(v) \tilde R_j(u) & = & \tilde R_j(u)
\tilde R_i(v) \quad \mbox{if}\ \ |i-j| \ge 2, \\
\tilde R_{i+1}(v-u) \tilde R_i (v) \tilde R_{i+1}(u) & = &
\tilde R_i(u) \tilde R_{i+1} (v) \tilde R_i(v - u).
\end{eqnarray}
Without the spectral parameters the $\tilde R$ matrices satisfy the
braiding relations. The complete description of the fractional
statistics based on the YBE needs to be worked out.

At present it seems unclear to me how the construction based on
the YBE can be applied to a general 1D (i.e., not necessarily
integrable) systems.  Maybe there is a way to map a class of
general 1D models to integrable models with in some sense small
residual interactions that break the integrability.
Instead of pursuing this incomplete Yang-Baxter story any further,
I give here somewhat heuristic but intuitive device for constructing the 1D
fractional statistics.
The $N$-particle state is constructed as follow
\begin{equation}
\int dx_1 \cdots dx_N \Psi(x_1,\ldots,x_N,t)
\phi^*(x_N,t)\cdots\phi^*(x_1,t)|0\rangle,
\label{state}
\end{equation}
where $\phi^*(x,t)$ is the anyon creation operator.  The ``anyon'' fields
satisfy the following equal time commutation relations
\begin{equation}
[\phi(x,t), \phi(y,t)]_\lambda = [\phi^*(x,t), \phi^*(y,t)]_\lambda =
[\phi(x,t), \phi^*(y,t)]_\lambda = 0 \mbox{if $x \ne y$},
\end{equation}
where $[x,y]_\lambda = xy + \exp(i\pi\lambda \mbox{sgn}(y-x)) yx$.

The wavefunction $\Psi(x_1,\ldots, x_N,t)$ is multi-valued and satisfies
\begin{equation}
\Psi(\ldots,x_i,\ldots,x_j\ldots,t) =
e^{i\pi\lambda\mbox{sgn}(x_i - x_j)} \Psi(\ldots,x_j,\ldots,x_i,\ldots,t).
\label{multiv}
\end{equation}
The integrand in Eq.\ (\ref{state}), however, is always single valued since
the phases arising from the wavefunction and the ``anyon'' field operators
are set to cancel each other.

If I specialize to the CSM the fractional exchange statistics can
be formulated in the first quantized language.
First, the Hamiltonian (\ref{hamil}) can be rewritten as \cite{poly}
\begin{equation}
H = {1\over 2m} \sum_{j=1}^N \left(p_j + i {\pi \hbar \lambda \over L}
\sum_{k(\ne j)} \cot\left[{\pi(x_j-x_k)\over L}\right]P_{jk}\right)^2,
\label{newhamil}
\end{equation}
where $p_j = -i\hbar \partial_{x_j}$ is the momentum operator and $P_{ij}$
the particle exchange operator.  The extra term added to the momentum
operator is an 1D analog of the Chern-Simons gauge field.

In two-dimension there are two well-known ways to code the fractional
statistics for the ideal anyon gas.  One is to take the free Hamiltonian
and require that its wavefunctions be multi-valued as in Eq. (\ref{multiv}).
The other is to introduce the Chern-Simons gauge field and write
the Hamiltonian in terms of this gauge field and further require
that the wavefunctions
be single valued and symmetric.  The Hamiltonian (\ref{newhamil})
corresponds to the 1D version of the second formulation of the
fractional exchange statistics.  Hence, the following symmetric
wavefunctions are the eigenstates of the 1D anyon system
\begin{equation}
\Psi^\lambda_{\mbox{\boldmath $\kappa$}} (x_1,\ldots, x_N) =
\varphi(x_1,\ldots,x_N) \Psi_0 J^\lambda_{\mbox{\boldmath $\kappa$}},
\label{anywave}
\end{equation}
where the ``ordering function'' $\varphi$ is introduced to maintain the
total wavefunction symmetric \cite{ha94}.  In particular $\varphi$ keeps
track of the braiding of the particles and is set to cancel the
exchange phases arising from $\Psi_0$. I require
the ordering function to satisfy
\begin{equation}
\varphi(x_1,\ldots,x_m,\ldots,x_N) = e^{i(m-1)\pi\lambda}
\varphi(x_1,\ldots,\hat x_m,\ldots, x_N),
\label{phase}
\end{equation}
where the hat over the variable denotes the absence of that variable from
the function.  In other words, to remove the particle at $x_m$ it is
necessary to pass through $m-1$ particles.

In Section \ref{sec:edge} I construct explicit second quantized
``anyon'' fields and show their consistency with the first
quantized formulation of the fractional exchange statistics for the
CSM.

\subsection{Exclusion Statistics}

The notion of fractional {\it exclusion} statistics based on the
so called ``generalized
Pauli exclusion principle'' has first been formulated by Haldane and applied
to the elementary topological excitations of general condensed matter systems
\cite{duncan92}.  This new concept of statistics is
based on the structure of the single particle Hilbert space of the
elementary excitations.  More specifically,
the change in the size of the available states ($\Delta D$) in the
Hilbert space as the number of particles (i.e., the elementary excitations)
is changed ($\Delta N$) for a given system with fixed boundary
condition defines
the statistics of the particles with the statistical
parameter defined as $g = - \Delta D/ \Delta N$.  Hence, for example,
the bosons and fermions are identified with $g = 0$ and $g = 1$,
respectively.

In order to facilitate proper understanding of the {\it exclusion}
statistics in the context of the CSM I introduce a pictorial
representation of the eigenstates and make the identification of
the excitation contents of the states easier \cite{ha94}.
Eq. (\ref{ks}) gives the occupation
configurations of the pseudomomenta $k_j$ for all the eigenstates of
the CSM.
The quantum numbers $\{I_j\}$ in Eq. (\ref{ks}) are distinct (half-odd)
integers and in the ground state are given by the following set
\begin{equation}
\{I_j^0\} = \left\{ -{N-1\over 2}, -{N-3\over 2}, \ldots, {N-3\over 2},
{N-1\over 2}\right\}.
\end{equation}
Therefore, the ground state pseudomomenta are given by
\begin{equation}
\{k_j^0\} = \left\{-{\pi\lambda\over L}(N-1), -{\pi\lambda\over L}(N-3),
\ldots, {\pi\lambda\over L}(N-3), {\pi\lambda\over L}(N-1)\right\}.
\end{equation}
The total ground state energy $E^0 = \sum_j (k_j^0)^2$ is equal to
$\pi^2\lambda^2 N(N^2-1)/3L^2$.
The excited states are given by integer displacements of $\{I_j^0\}$.
Therefore, two neighboring pseudomomenta for any arbitrary state
must be separated by
\begin{equation}
\Delta k_j \equiv |k_j - k_{j-1}| = {2\pi\over L}(\lambda + l),
\end{equation}
where $l$ is a non-negative integer.

In order to construct a picture that exposes the excitation content of
the excited states, I let $\lambda$ to be a rational number $p/q$ with
$p$ and $q$ coprimes and introduce one-dimensional lattice with the
lattice spacing equal to $2\pi/qL$. I assign each lattice point with
$1$ if that lattice point coincides with the value of one of the
occupied pseudomomenta and with $0$ if it does not. Hence,
the ground state for $\lambda = 3/2$ and $N = 10$ is represented by
$\ldots 0000000010010010010010010010010010010000000\ldots$.  All the other
excited states can be obtained from this ground state configuration
by displacing the {\it ones} such that the number of {\it zeroes}
between any pair of {\it ones} is equal to $p - 1 + q l$ where
$l$ is a non-negative integer.

I use three different names for the particles in
the model---real, pseudo, and quasiparticles.
The real particles are, of course, the
physical quantum particles described by the canonically conjugate
coordinate and momentum variables $\{x_j, p_j\}$.  The pseudoparticles
are described by the pseudomomentum operator (see Eq.\ (\ref{newhamil}))
which a sum of the usual momentum and the 1D ``statistical gauge field.''
The quasiparticles are the elementary excitations of the
system. Because the pseudoparticles form an ideal gas,
the quasiparticles are essentially same as the pseudoparticles excited
out of the condensate.  The holes left behind in the pseudoparticle
condensate will be called quasiholes.  Hence, the name ``pseudo''
and ``quasi'' will be used interchangeably in some cases \cite{ha94}.

Since $p-1$ {\it zeroes} are always required between the {\it ones}
I call them {\it bound zeroes} which seem to represent the
mutual statistical exclusions.  In the case of the free Fermions
carrying the flux $2\pi$ in units where $e = \hbar = c = 1$ so that
the flux quanta $\Phi_0 = 2\pi$,
the minimum separation of the momenta is $2\pi/L$ which can be considered
as the mutual Pauli exclusion. In this case the minimum separation
is $2\pi\lambda$; therefore, it is natural to assign $2\pi\lambda$
flux attached to the pseudoparticles.
The remaining {\it zeroes} are
called {\it unbound zeroes}. The $q$ consecutive unbound {\it zeroes}
in the condensate of the pseudoparticles constitute a
single hole excitation.  Thus, if a pseudoparticle is removed from the
ground state condensate, then there are $p$ {\it unbound zeroes} in place
where the {\it one} is removed.  This state is forbidden if
$q \ne 1$. In general a minimum of $q$ {\it ones} must
be removed so that they leave behind at least $pq$ {\it unbound zeroes}
which break up into $p$ holes.  From the view point of the particles (holes)
the change in the number of available single particle states is
$p$ ($-q$) while the change in the number of quasiparticles
(quasiholes) in the system is
$-q$ ($p$). Therefore, the statistical parameter $g$ for the
quasiparticle (quasihole) is $g = p/q = \lambda$ ($g = q/p = 1/\lambda$).
This inverse relationship between the statistical parameters of the
particles and holes is essentially the Chern-Simons duality.
To summarize the fractional exclusion statistics, {\it q particle
excitations are accompanied by p hole excitations.}

The configurations constructed above are representations of
the diagrams of partitions ${\cal D}({\mbox{\boldmath $\kappa$}})$
introduced in Sec. \ref{sec:Jack}.
The part $\kappa_j$ corresponds to the displacement of $j$th quantum number
from the ground state (i.e., $I_j - I_j^0$ if $I_1 > I_2 > \ldots > I_N$).
The excitations given by {\boldmath$\kappa$} include only the states
with non-negative displacements (i.e., $k_j$ moved only to the right) and
all the other states are obtained by global Galilean transformations.
Therefore, each row (column) in the diagram corresponds to the
particle (hole) excitations of the CSM.

Following Yang and Yang \cite{yy} and Sutherland \cite{suther}
I can also construct the thermodynamics \cite{ha94,denise94}.
In the thermodynamic limit it is convenient to define
the hole distribution $\rho_h(k)$ and the particle distribution
function $\rho_p(k)$.  The function $\rho_p(k)$ are given by the solutions
of Eq. (\ref{ks}) while $\rho_h(k)$ by the corresponding complementary
equations given by the unused quantum numbers $I_j$; thus, they satisfy
$ 1 = \rho_h(k) + \lambda \rho_p(k)$.  This equation states that
one hole and $1/\lambda$ particles (or
$\lambda$ holes and one particle) have equal weight in
occupying the volume in $k$-space.
Hence, for $\lambda \ne 1$ the particle-hole
symmetry is broken; and in the bosonic case ($\lambda = 0$) the
symmetry is maximally broken.  With a proper normalization one can equally
state that $1 = (1/\lambda) \rho_h(k) + \rho_p(k)$.  This particle-hole
duality is the essence of the Chern-Simons duality.

The thermodynamic function is given
by  $\Omega = E - TS - \mu N$, where the energy ($E$), the particle
number (N), and the entropy ($S$) are given by
\begin{eqnarray}
E & = & V \int dk \rho_p(k) e(k), \\
N & = & V \int dk \rho_p(k), \\
S & = & V \int dk \{(\rho_h + \rho_p)
\log(\rho_h+\rho_p)-\rho_h\log\rho_h-\rho_p\log\rho_p \}.
\end{eqnarray}
Here, $V$ is the volume of the system.  By minimizing $\Omega (\rho_p,
\rho_h)$ with respect to the density functions the following relation
can easily be obtained
\begin{equation}
(1-\lambda \rho_p(k))^\lambda (1+(1-\lambda)\rho_p(k))^{1-\lambda}
 = \rho_p(k) e^{(e(k) - \mu)/T}.
\end{equation}
This equation is also obtained by Wu using a different method
\cite{wu94}.

\section{Dynamical density-density correlation function}
\label{sec:ddcf}

In this section I show that the dynamical density-density correlation
function (DDDCF) can be calculated exactly for the CSM using the
known properties of Jack polynomials.
The DDDCF is defined by
\begin{eqnarray}
\mbox{}_N\langle 0|\rho(x',t')\rho(x,t)|0\rangle_N & = &
\mbox{}_N\langle 0|e^{iH_Nt'}\rho(x')e^{-iH_Nt'}e^{iH_Nt}
\rho(x)e^{-iH_Nt}|0\rangle_N \nonumber \\
\mbox{} & = &
\mbox{}_N\langle 0|\rho(x')e^{-i(H_N-E_N^0)(t'-t)}\rho(x)|0\rangle_N,
\end{eqnarray}
where the {\it reduced} density operator  $\rho(x) = (1/L) \sum_j
\delta(x-x_j) - N/L$ and $|0\rangle_N$ the normalized $N$-particle
ground state.

The first step in calculating the DDDCF is to
expand $\rho(x)|0\rangle_N$ in terms of the eigenstates of the CSM
(i.e., Jack polynomials).  The delta function $\delta(x)$ is a
periodic function with a period $L$ and thus can be expressed as
a Fourier sum $(1/L)\sum_{m=-\infty}^{+\infty} \exp(i2\pi x m/L)$.
Therefore, I can write $\rho(x)$ as follow
\begin{equation}
\rho(x) = {1\over L} \sum_{m = 1}^\infty (z^m p_{-m} + z^{-m} p_m),
\end{equation}
where $z = \exp(i2\pi x/L)$, $z_j = \exp(i2\pi x_j/L)$,
and $p_m = \sum_{j=1}^N z_j^m$. The power sum $p_m$ can be
expanded in terms of Jack polynomials using the
identity Eq. (\ref{power}).

Using the orthogonality relation Eq. (\ref{orth}) and
its extension Eq. (\ref{orth1}), I obtain the
following expression for the DDDCF
\begin{equation}
\mbox{}_N\langle 0|\rho(x,t)\rho(0,0)|0\rangle_N = {1\over L^2}
{2\over \lambda^2}\sum_{\mbox{\boldmath $\kappa$}}
{|{\mbox{\boldmath $\kappa$}}|^2 \over j_{\mbox{\boldmath $\kappa$}}^\lambda}
{([0']_{\mbox{\boldmath $\kappa$}}^\lambda)^2
[N]_{\mbox{\boldmath $\kappa$}}^\lambda \over
[N+1/\lambda - 1]_{\mbox{\boldmath $\kappa$}}^\lambda}
\cos(2\pi|{\mbox{\boldmath $\kappa$}}|x/L)
e^{-itE_{\mbox{\boldmath $\kappa$}}},
\label{finited}
\end{equation}
where $E_{\mbox{\boldmath $\kappa$}}=(2\pi/L)^2 \sum_{j=1}^N (\kappa_j^2 +
\lambda (N+1-2j)\kappa_j)$.
The coefficient $[0']_{\mbox{\boldmath $\kappa$}}^\lambda$
in Eq. (\ref{finited}) vanishes
unless the diagram ${\cal D}({\mbox{\boldmath $\kappa$}})$ has
no more than $p$ columns of length longer than q and $q$ rows
of length longer than p.  In other words, the
intermediate states contributing to the DDDCF
has precisely $p$ hole and $q$ particle excitations.
This is a conclusive evidence of the ideal fractional exclusion
statistics the CSM quasiparticles and quasiholes obey.

The DDDCF in the thermodynamic limit greatly simplifies as is shown in
Appendix A; and it is given by \cite{ha94}
\begin{equation}
\langle0|\rho(x,t)\rho(0,0)|0\rangle = C\ \prod_{i=1}^q
\left(\int_0^\infty dx_i \right)
\prod_{j=1}^p \left(\int_0^1 dy_j\right) Q^2 F(q,p,\lambda|\{x_i,y_j\})
\cos(Qx) e^{-iE t},
\label{ddc}
\end{equation}
where $Q$ and $E$, the total momentum and energy, are given in units
of $\hbar$ and $\hbar^2/2m$ by
\begin{eqnarray}
Q & = & 2\pi\rho_0\left(\sum_{j=1}^q x_j + \sum_{j=1}^p y_j\right),
\label{momen}\\
E & = & (2\pi\rho_0)^2\left(\sum_{j=1}^q \epsilon_P(x_j) +
\sum_{j=1}^p \epsilon_H(y_j)\right), \label{energ}
\end{eqnarray}
with $\rho_0 = N/L$, $\epsilon_P(x) = x(x+\lambda)$ and
$\epsilon_H(y) = \lambda y(1-y)$.
$x_j(\epsilon_P)$ and $y_j(\epsilon_H)$
are normalized momentum (energy) of
the quasiparticles and the quasiholes, respectively.
The normalization constant $C$ is given by
\begin{eqnarray}
A(m,n,\lambda) & = & {\Gamma^m(\lambda) \Gamma^n(1/\lambda)
\over \prod_{i=1}^m \Gamma^2(p-\lambda(i-1))
\prod_{j=1}^n \Gamma^2(q-(j-1)/\lambda)} \prod_{j=1}^n
{\left(\Gamma(q-(j-1)/\lambda) \over \Gamma(1-(j-1)/\lambda)\right)}^2, \\
C & = & {\lambda^{2p(q-1)} \Gamma^2(p)\over 2\pi^2 p!q!} A(q,p,\lambda).
\label{norm}
\end{eqnarray}
Finally, the form factor $F(q,p,\lambda|\{x_i,y_j\})$ is given by
\begin{equation}
F(m,n,\lambda|\{x_i,y_j\}) =
\prod_{i=1}^m \prod_{j=1}^n (x_i + \lambda y_j)^{-2}
{\left(\prod_{i<j} (x_i - x_j)^2\right)^\lambda
\left(\prod_{i<j} (y_i - y_j)^2\right)^{1/\lambda}
\over
\prod_{i=1}^m \epsilon_P(x_i)^{1-\lambda}
\prod_{j=1}^n \epsilon_H(y_j)^{1-1/\lambda}}.
\label{formfac}
\end{equation}
The form factor has been conjectured by Haldane based on the clues
given by the works of Simons {\it et. al.}, Galilean invariance, and
$U(1)$ conformal field theory \cite{duncan942}.  A less general form of
the DDDCF at integer values of $\lambda$ has also been reported
\cite{other}.

The region of support in the energy-momentum space
for the DDDCF at $\lambda = 5/3$
corresponds to the shaded area in Fig. \ref{sector}.  It is
obtained by convoluting the dispersion relations of the $q$ quasiparticles
and $p$ quasiholes as given by Eqs. (\ref{momen}) and (\ref{energ}).
At low-energies $2p$ distinct sectors indicated by darker shade
emerge as expected of 1D metallic system. (A generic 1D system, however,
will have an infinite number of these low-energy sectors.)

It is also worth noting that there is a qualitative difference
between the finite and the infinite system.
The intermediate states represented by the diagrams
with $|\kappa_j - \kappa_k|$ and $|\kappa_j' - \kappa_k'|$
of order ${\cal O}(N)$ are the only states that
contribute to the DDDCF in the thermodynamic limit
(i.e., $N,\ L \rightarrow \infty$ with $N/L$ fixed).
I call this a super-selection rule \cite{ha94}.  (There can be
violations of this rule at the low-energy limit as discussed in Appendix A.)
Perhaps, the phase transition that occurs in 2D QCD as $N \rightarrow \infty$
\cite{qcd} is related to the CSM super-selection rule.

\section{One-Particle Green's Function}
\label{sec:gf}

Unlike the DDDCF calculated in the previous section, evaluation
of the one particle Green's function depends on the actual statistics of
the real particles since it involves the reordering of the
particles.  For example, the one-particle density matrix
(the static limit of the Green's function) for the $\lambda = 1$ case
is trivially simple if the ground state
$\Psi_0$ is taken to be the fermionic wave
function. On the other hand, if the wavefunction is taken to describe
a bosonic system, the calculation gets quite complicated and different
from its fermionic counterpart as shown in \cite{jj}.

As described in section \ref{sec:fractional} I take the
modified CSM given by Eq. (\ref{newhamil}) as
one-dimensional anyon system with the exchange
phase given by $\exp(i\lambda\pi)$ and
Eq. (\ref{anywave}) as the corresponding wavefunctions.
If the $m$th particle is removed from the ground state of
$N+1$ particles, the remaining term after
factoring out the ground state wavefunction
of the $N$ particle system is
\begin{equation}
\left(\prod_{j(<m)}(z_j-z_m)^\lambda \right)
\left( \prod_{j(>m)} (z_m-z_j)^\lambda \right)
\left( \prod_{k\ne m} z_k^{-\lambda/2} \right) z_m^{-\lambda N/2}
e^{i(m-1)\pi\lambda},
\end{equation}
where the last term is determined from the property of the ordering
function $\varphi$ given in Eq. (\ref{phase}).
Therefore, up to an overall phase factor that does not depend on the position
of the removed particle, the destruction operation $\Psi(x)$ on the
ground state of $N+1$ particles is given by
\begin{equation}
\Psi(x) |0\rangle_{N+1} = {A_{N+1}\over A_N} z^{-\lambda N/2}
\prod_{j=1}^N (z - z_j)^\lambda z_j^{-\lambda/2} |0\rangle_N,
\end{equation}
where $z = \exp(i2\pi x/L)$ and $A_N^2 = (1/L)^N\Gamma^N(1+\lambda)
/\Gamma(1+\lambda N)$.  The statistical phase arising from the
ordering function makes the destruction operation symmetric with respect
to the permutations of $\{z_j\}$.  The symmetric
function $\prod_{j=1}^N (z - z_j)^\lambda$ can then be expanded in term
of Jack polynomials using Eqs. (\ref{destruct}) and (\ref{homo}).

The hole propagator part of the one-particle Green's function is
defined as
\begin{eqnarray}
\mbox{}_{N+1}\langle0| \Psi^\dagger(x',t') \Psi(x,t) |0\rangle_{N+1}
& = & \mbox{}_{N+1}\langle0| e^{iH_{N+1}t'} \Psi^\dagger(x')
e^{-iH_Nt'} e^{iH_Nt} \Psi(x) e^{-iH_{N+1}t} |0\rangle_{N+1} \nonumber \\
& = & \mbox{}_{N+1}\langle0| \Psi^\dagger(x') e^{-i((H_N-E_N)-\mu)(t'-t)}
\Psi(x)|0\rangle_{N+1},
\end{eqnarray}
where the chemical potential $\mu = E_{N+1}-E_N$.
By expanding $\Psi(x)$ in terms of Jack polynomials and using the
orthogonality relation (\ref{orth})
the propagator at finite $N$ and $L$ is evaluated to be
\begin{eqnarray}
\mbox{}_{N+1}\langle 0| \Psi^\dagger(x,t) \Psi(0,0) |0\rangle_{N+1}
& = &(N+1) \left({A_{N+1}\over A_N}\right)^2 \nonumber \\
&\times& \sum_{\mbox{\boldmath $\kappa$}}
{\lambda^{-2|{\mbox{\boldmath $\kappa$}}|} \over
j_{\mbox{\boldmath $\kappa$}}^\lambda }
{(\{-\lambda\}_{\mbox{\boldmath $\kappa$}}^\lambda)^2
[N]_{\mbox{\boldmath $\kappa$}}^\lambda \over
[N+1/\lambda-1]_{\mbox{\boldmath $\kappa$}}^\lambda }
e^{i2\pi (|{\mbox{\boldmath $\kappa$}}|-\lambda N/2) x/L}
e^{-i(E_{\mbox{\boldmath $\kappa$}}-\mu)t},
\label{hp}
\end{eqnarray}
where the additional $(N+1)$ factor comes from the freedom
of choosing one of $N+1$ available particles to destroy and create.
The coefficient $\{-\lambda\}^\lambda_{\mbox{\boldmath $\kappa$}}$
vanishes unless the diagram ${\cal D}({\mbox{\boldmath $\kappa$}})$
has at most $q-1$ rows of length greater than $p$ and
$p$ columns of length greater than $q-1$.  Therefore, the intermediate
states for the propagator is spanned by $q-1$ quasiparticles
and $p$ quasiholes.  This is a very important result. {\it The exclusion
statistics of the quasiparticles and quasiholes is completely
consistent with the anyon statistics of the real particles.} \cite{ha94}

Taking the thermodynamic limit of Eq. (\ref{hp}) is almost identical
to that of Eq. (\ref{finited}) and is given by \cite{ha94}
\begin{equation}
\langle 0| \Psi^\dagger(x,t) \Psi(0,0) |0\rangle =
\rho_0 D \prod_{i=1}^{q-1}\!\!
\left(\int_0^\infty\!\! dx_i\right)\!\!
\prod_{j=1}^p \left(\int_0^1\!\! dy_j\right)
F(q-1,p,\lambda|\{x_i,y_j\}) e^{i((Q-Q_0)x-(E-\mu)t)},
\label{propagator}
\end{equation}
where the chemical potential $\mu = (\pi\lambda\rho_0)^2$ and
the back flow $Q_0 = \pi\lambda\rho_0$.
$F(q-1,p,\lambda|\{x_i,y_j\})$ is still given by Eq.\ (\ref{formfac})
and D by
\begin{equation}
D = {\lambda^{2p(q-1)} \Gamma^2(p) \over
\Gamma(\lambda) (q-1)!p!} A(q-1,p,\lambda).
\end{equation}
$Q$ and $E$ are same as before except for the number of $x_j$'s.
At integer values of $\lambda$ (i.e., $q=1$ case where only quasiholes
are excited), based on the equal-time results of Forrester \cite{forrester}
Haldane made a conjecture \cite{duncan94} which agrees with this formula.
The regions of support for the hole propagator is given by the
shaded area in $Q > 0$ (or $Q < 0$) in Fig. \ref{sector}.
There are also shifts in $E$ and $Q$ by $-\mu$ and $-Q_0$, respectively.

It is also interesting to consider the following function
\begin{equation}
\Psi_m^\lambda(x) = \prod_{i<j} (z-z_j)^m,
\label{mpsi}
\end{equation}
where $m$ is a positive integer.  This function can be expanded
in terms of Jack polynomials using Eq. (\ref{destruct}) with
coefficients containing the term
$\{-m\}^\lambda_{\mbox{\boldmath $\kappa$}}$ which vanishes unless
the diagram ${\cal D}(\mbox{\boldmath $\kappa$})$ has no more than $m$
columns and no rows longer than $m$.
Therefore, $\Psi_m^\lambda(x)$
acting on the ground state creates exactly $m$ quasiholes.
$\Psi_m^\lambda(x)$ is a generalization
of the one-hole state given in Eq. (\ref{onehole}).
The propagator can easily be calculated and its form factor is given by
$F(0,m,\lambda|\{y_j\})$.  This result is consistent with the conjecture
\cite{ha94} that the minimal form factor for any two point
correlation functions whose intermediate states contain only
$n$ quasiparticle and $m$ quasiholes is given by
$F(n,m,\lambda|\{x_i,y_j\})$.

\section{Low-energy effective theory and
Coupled Fractional Quantum Hall Edge States}
\label{sec:edge}

Following the standard method in Luttinger liquid theory \cite{lutt}
I construct an effective low-energy model for the 1D anyon system.
When $\lambda$ is an integer this system is equivalent to a coupled
system of left- and right-moving edge states of the FQHE.
The excitations on a single edge of the FQH fluid moves
only in one direction because of
the externally applied magnetic field; and they have
been rather thoroughly studied using the so called
chiral Luttinger liquid theory \cite{edge} which is
intrinsically anomalous \cite{chiral}.
When there is an extra edge on the FQH droplet (e.g. a strip, annulus,
or cylindrical geometry instead of the disk geometry) and when the
edges are close enough, new phase space
opens up as a result of the fractional charge transfer between the edges.
There actually have been many suggestions that the CSM is related to the
edge states \cite{csmedge}.  In this section I show in another way
that the CSM is an exactly solvable 1D anyon system by
calculating the correlation functions of the effective model
and finding exact agreement with those of the CSM
in the long-wavelength limit.

A general gauge invariance argument \cite{gauge}
can be used to map out the qualitative
structure of the excitation spectra in the energy-momentum space.
I use the cylindrical geometry \cite{gauge} with coordinates $x$
and $y$ defined as shown in Fig. \ref{cylinder}.
I put some incompressible fluid perhaps made up of $\lambda$ anyons
on the surface of the cylinder with a confining potential in
$y$-direction such that two edges are created along $x$-direction.
(Of course, when $\lambda$ is not an integer the edges are not related
to the FQHE edges which are more complicated composite type \cite{edge}.)
The cylinder with circumference $L$ is also pierced by a thin
solenoid along its longitudinal axis, thus inducing a magnetic flux
$\Phi = AL$ through its cross-section and a vector potential
$\vec A = A \hat x$ on its surface.
There is also magnetic fields $\vec B$ normal to
the entire surface of the cylinder.
Since the left- and right-moving sectors of the CSM completely decouple
in the low-energy limit, this geometrical construction is in fact
equivalent to the CSM and is useful to show intuitively how the low-energy
sectors are related.

The fictitious flux applied to the
cylinder is used as a passive device for mapping out the regions of
low-energy excitation sectors of the coupled anyon edges.
By virtue of the gauge invariance the energy is degenerate at
flux equal to $2\pi n$ (in units where $\hbar = e = 1$) which
corresponds to the momentum of $2\pi n \rho_0$.  The adiabatical
change in flux from $0$ to $2\pi$ induces an elementary excitation
carrying charge $-q/p$ and flux $-2\pi$ (i.e., a quasihole)
to move from one edge to the other; and, in between,
it costs finite energy to the system since the bulk is incompressible.
In fact the real space configuration of an incompressible droplet
is equivalent to a ``Fermi sea.'' \cite{fradkin}
The edges of the fluid coincide with the locus of points in which
Fermi energy crosses the external confining potential.
The only degrees of freedom left in ``Fermi sea'' in the low-energy limit
is the ``Fermi surface'' fluctuation.
This ``Fermi sea'' obeys the fractional exclusion statistics.
Since there is a symmetry along $x$-direction, the fluid can be regarded
as an 1D ``Fermi sea'' and the left- and the right-edges as two
``Fermi points.''

In light of the observations made so far it is reasonable
to model the coupled edges by a full (non-chiral) 1D anyon system.
First, I construct the one-dimensional anyon creation operator as follow
\begin{equation}
\Psi^\dagger_\lambda(x) = \Psi^\dagger_B(x) e^{i\lambda \theta(x)},
\end{equation}
where $\Psi^\dagger_B$ is the boson creation operator and $\lambda$ the
statistical parameter. The operator
$e^{i\lambda \theta(x)}$ is the so called ``disorder operator''
\cite{disorder} that creates a kink (or vortex) of size $\pi \lambda$
at position $x$ and is defined as
\begin{equation}
 \theta(x) = \pi \int_{-\infty}^x \rho(x') dx',
\end{equation}
where $\rho(x)$ is the density operator.
Therefore, $\Psi^\dagger_\lambda(x)$ is a composite
operator that creates a boson plus a vortex; so, it is an anyon
creation operator.

The boson creation operator $\Psi^\dagger_B(x)$ for a system with
multiple low-energy sectors has previously been constructed
\cite{duncan82} and is given by
\begin{equation}
\Psi^\dagger_B(x) \approx \rho_0^{1/2} \sum_{m=-\infty}^{+\infty}
e^{i2m \theta(x)} e^{i\phi(x)},
\end{equation}
where the phase field $\phi(x)$ is defined by the following
canonical commutation relation: $[\phi(x), \rho(x')] = i\delta(x-x')$.
The multi-sector density operator is also given by
\begin{equation}
\tilde \rho(x) = \rho(x) \sum_{m=-\infty}^{+\infty} e^{i2m\theta(x)}.
\end{equation}
Here, the sectors are connected by the operator $\exp(i2m\theta(x))$ which
creates a vortex of size $2\pi m$.  This dynamical device now
replaces the passive device I previously used to map the system
from one sector to the next by supplying external magnetic flux
to the cylinder.

The fields $\theta(x)$, $\phi(x)$ and $\rho(x)$ can now be expressed
in terms of Tomonaga boson operators $b^\dagger$ and $b$
as \cite{duncan82,mahan}
\begin{eqnarray}
\theta(x) & = & \theta_0 + \pi \rho_0 x + i e^{\varphi}
\sum_{k\ne 0} {1\over k} \left|{\pi k\over 2 L}\right|^{1/2}
e^{-ikx}(b^\dagger_{-k} + b_k) \\
\phi(x) & = & \phi_0 - i e^{-\varphi} \sum_{k\ne 0}
\left|{\pi \over 2k L}\right|^{1/2} e^{-ikx}(b^\dagger_{-k} - b_k) \\
\rho(x) & = & \rho_0 + e^\varphi \sum_{k\ne 0}
\left|{k\over 2\pi L}\right|^{1/2} e^{-ikx}(b^\dagger_{-k} + b_k).
\end{eqnarray}
I assume here that the most important interaction is
the statistical interaction; so, I can set the Bogoliubov
parameter $e^{2\varphi} = 1/\lambda$.

Now, $\Psi^\dagger_\lambda(x)$ can easily be shown to satisfy the following
anyon commutation relation
\begin{equation}
\Psi^\dagger_\lambda(x) \Psi^\dagger_\lambda(x') =
e^{i\pi\lambda \mbox{sgn}(x'-x)}
\Psi^\dagger_\lambda(x')\Psi^\dagger_\lambda(x)
\quad \mbox{for $x \ne x'$}.
\label{comm}
\end{equation}
Eq. (\ref{comm}) can thus be used as a defining relation for the
one-dimensional anyons.

The Hamiltonian is diagonal in Tomonaga boson operators and is
given as usual by
\begin{equation}
H = \hbar v_s \sum_{q\ne 0} |q| b^\dagger_q b_q,
\end{equation}
where $v_s$ is the sound velocity.

Using the operator identities (i) $e^A e^B = \exp(e^\alpha -1)e^B e^A$
if $[A, B] = \alpha B$, (ii) $e^A e^B = \exp(-[A,B])e^B e^A$
and $e^{A+B} = \exp(-[A,B]/2) e^A e^B$ if $[[A,B],A]
=[[A,B],B] = 0$, the following correlation functions are calculated
\begin{eqnarray}
\langle \hat \rho(x,t) \hat \rho(0,0)\rangle & \approx & \rho_0^2
\left(1 - {\lambda^{-1} \over (2\pi\rho_0)^2}
     \left({1\over (\xi^-)^2} + {1\over (\xi^+)^2}\right) \right.\nonumber \\
     & + &\left. \sum_{m=1}^\infty A_m \left({1\over \xi^+\xi^-}
     \right)^{m^2/\lambda} \cos(2\pi\rho_0 m x)
\right), \label{dd} \\
\langle\Psi^\dagger_\lambda(x,t) \Psi_\lambda(0,0)\rangle & \approx & \rho_0
\sum_{m=-\infty}^\infty B_m \left({1\over \xi^+}\right)^
{(m+\lambda)^2/\lambda} \left({1\over \xi^-}\right)^{m^2/\lambda}
e^{i(2\pi\rho_0(m+\lambda/2)x + \mu t)} \label{gf},
\end{eqnarray}
where $\xi^\pm = x \mp v_s t$, $\mu$ the chemical potential
and the coefficients $A_m$ ($B_m$) are regularization-dependent constants.

The Green's function
$\langle\Psi^\dagger_\lambda(x,t) \Psi_\lambda(0,0)\rangle$ for the
sector $m=0$, where the charge transfer from one edge to the other
is forbidden, is given by only the right-movers (or only the left-movers
if $\Psi^\dagger_\lambda$ were properly redefined) even though
the anyon creation operator
$\Psi^\dagger_\lambda(x)$ contains both the right- and the left-moving
bosonic modes (i.e., $b^\dagger_k$ and $b_k$ for both $k > 0$ and
$k < 0$).  This chiral sector emerges
naturally in this theory without explicitly imposing the
chirality condition.

In an {\it isolated} chiral theory coupled to a gauge field
the charge is in general not conserved and
because of this the theory is known to be
anomalous and not physical.  As far as the isolated
chiral theory is concerned the
only physical sector is the chiral sector where no charge transfer between
the edges is possible.  One can, however, consider
an ``almost'' chiral theory which describes a
system of two chiral edge states that are
independent except for the charge leakage from one side
to the other and vice versa. For the sake of definiteness I concentrate
on the right edge and propose the ``almost'' chiral system with
the following Hamiltonian and the field operators,
\begin{eqnarray}
H^R & = & \hbar v_s \sum_{q>0} q b^\dagger_q b_q, \\
\theta^R(x)& = &\theta_0 + \pi \rho^R_0 x - {i\over \sqrt{\lambda}}
\sum_{k>0} {1\over k} \left|{\pi k\over 2L}\right|^{1/2}
\left(e^{ikx}b^\dagger_k - e^{-ikx} b_k\right), \\
\phi^R(x) & = &\phi_0 -i \sqrt{\lambda} \sum_{k>0}
\left|{\pi \over 2kL}\right|^{1/2} \left( e^{ikx} b^\dagger_k
- e^{-ikx} b_k\right), \\
{\Psi^R_\lambda}^\dagger(x) &\approx& \sqrt{\rho_0^R}
\sum_m e^{i2(m+\lambda/2)\theta^R(x)} e^{i\phi^R(x)}.
\end{eqnarray}
The chiral system is constructed only with the right-moving
Tomonaga bosons.  The Green's function in this case is given by
\begin{equation}
\langle {\Psi^R_\lambda}^\dagger (x,t) \Psi^R_\lambda(0,0)
\approx \rho^R_0 \sum_{m = -\infty}^\infty C_m \left({1\over
x - v_s t}\right)^{(m+\lambda)/\lambda} e^{i(2\pi\rho^R_0(m+\lambda/2)x +
\mu^R t)}.
\end{equation}
The $m = 0$ sector here is equivalent to the corresponding
chiral sector of the non-chiral model.  As expected only the
right-movers contribute to the Green's function for all the
other sectors.

Asymptotic expansions of the correlation functions of the CSM
have been calculated in Appendix B; and they agree with
Eq. (\ref{dd}) and Eq. (\ref{gf}).
This agreement between the correlation functions of the explicitly
constructed anyon model and the CSM shows
that the first- and the second-quantized
construction of the anyons are in fact equivalent in the long-wavelength
limit.

\section{Finite-size Scaling and correlation exponents}
\label{sec:finite}

There is another extremely elegant and powerful way of obtaining the
exponents of the correlation functions in the long-wavelength limit.
If the dispersion relations of the elementary excitations
of the one-dimensional quantum model in the low-energy limit
is linear (i.e., has Lorentz as well as Galilean invariance so that
the space and time variables are on equal footing),
the principle of conformal invariance is applicable and,
in particular, the finite-size corrections to the energy and
momentum become universal and are directly related to the
exponents of the correlation functions \cite{cardy}.

The finite-size scaling has previously been
applied to the CSM by Kawakami and
Yang \cite{kawayang} without the benefit of
recently uncovered knowledge \cite{duncan94,ha94}, namely that due to
the ideal fractional statistics the CSM particles obey
the intermediate states for the DDDCF and the Green's function
are spanned by finite number of the elementary excitations.
The shaded regions in Fig. \ref{sector} are
the relevant sectors for the correlation functions
(for the Green's function only the positive or negative momentum sectors
contribute).
A complete set of relevant exponents is obtained
using the finite-size scaling in this section.

Let $(p_L, h_L | p_R, h_R)$ be a label for a low-energy sector
spanned by $p_L$ ($h_L$) left-moving
and $p_R$ ($h_R$) right-moving quasiparticles (quasiholes).
The relevant sectors for the DDDCF
are $(0,n|q,p-n)$ and $(q,p-n|0,n)$ and
for the Green's function $(0,n|q,p-n)$ or $(q,p-n|0,n)$ where
$n = 0, 1, \ldots, p$.
First, for the DDDCF the
lowest-energy state in the sector $m$ is characterized by $m$
quasiholes on right (or left) side of the ``Fermi sea'' transfered
to the left (or right).  Hence, the pseudomomenta $k_j$ for this
sector is given in terms of the ground-state pseudomomenta $k_j^0$
as $k_j = k_j^0 + 2\pi m/L$; and the energy and momentum by
\begin{eqnarray}
E_m & = & E_0 + {2\pi v_s\over L} {m^2\over \lambda}, \\
P_m & = & 2\pi m \rho_0,
\end{eqnarray}
where the sound velocity $v_s = 2\pi\lambda\rho_0$.  The scaling dimension
$x = h^+ + h^-$ of the density operator and its conformal spin
$s = h^+ - h^-$ are therefore given by $x = m^2/ \lambda$ and $s = 0$,
where $h^\pm$ are right (left) conformal weights and
$2h^\pm$ are the exponents that actually appear in the correlation
functions for the right(left) movers.
Thus, $2h^+ = 2h^- = m^2/\lambda$ in agreement with the exponents found
in the previous section and in Appendix B.

Now, for the hole propagator the energy and momentum of the sector $m$
characterized by $m$ quasihole transfers from the left to
right ``Fermi points'' and thus the
pseudomomenta $k_j = k_j^0 + 2\pi m/L$, where $j = 1,\ldots, N-1$, are
\begin{eqnarray}
E_m & = & E_0 - \mu + {2\pi v_s \over L}\left({m^2\over \lambda} - n +
{\lambda \over 2}\right), \\
P_m & = & 2\pi (n - \lambda/2) \rho_0 + {2\pi\over L} \left(
{\lambda \over 2} - n \right).
\end{eqnarray}
The chemical potential $\mu = (\pi\lambda\rho_0)^2$ is associated with
the particle destruction.  I remove the rightmost pseudoparticle
with $k_N^0 = \pi\lambda(N-1)/L$ from the ground state condensate and
do not introduce any separate selection rules in contrast to Ref.
\cite{kawayang}.
In this case the right and left conformal
weights are different (due to the non-zero conformal spin
$s = \lambda/2 - n$)
and are given by $2h^+ = (m-\lambda)^2/\lambda$ and
$2h^- = m^2/\lambda$ as expected. The right and left conformal
weights would be switched if the excitations were caused by the removal
of the leftmost pseudoparticle at $-\pi\lambda(N-1)/L$.

Apparently, the long-range interaction does not destroy the conformal
invariance of the CSM in the long-wavelength limit.  This is expected
from the linear dispersion relations of the CSM quasiparticles and
quasiholes in this limit.

\section{Lattice Models}
\label{sec:lattice}

The Haldane-Shastry model (HSM) corresponds to a lattice generalization
of the CSM at $\lambda = 2$ and has the following Hamiltonian
\begin{equation}
H = J_0 \sum_{i<j} {{\vec S_i}\cdot{\vec S_j}\over d^2(i-j)},
\end{equation}
where $d(m) = (N/\pi)|\sin(\pi m/N)|$ and $\vec S_j$ is the $SU(2)$ spin
operator acting on site $j$.
The model possesses a quantum symmetry called Yangian \cite{yangian} and
exhibits supermultiplets structures whose spin contents
are exactly reproducible from an asymptotic limit of the
thermodynamic Bethe-ansatz equations
\cite{string}.  Furthermore, there is one-to-one correspondence
between the highest weight states of the Yangian supermultiplets and the
states of the CSM (i.e., they satisfy the same
eigenvalue equations.)  A further generalization to $SU(n)$ case
has also been accomplished in \cite{hahal92,kawa}.
The lattice CSM model at even (odd) integer values of $\lambda$ can
be mapped to bosonic (fermionic) spinless t-J model \cite{hahal92}.

The method for evaluating the correlation functions of the CSM
presented in this paper is directly applicable to its lattice cousins
in some limited cases.
The Galilean invariance is broken in the lattice models and is
replaced with much weaker lattice translation symmetry which
induces appearance of the Brillouin zone boundaries.
If the elementary excitations created by a local operator acting
on the ground state
do not cross the Brillouin zone boundaries and the
excited states are the Yangian highest weight states,
then the corresponding correlation functions of HSM are identical to
that of the CSM.

\section{Conclusion}

In this paper the fractional exchange and exclusion
statistics are studied using the exactly solvable
Calogero-Sutherland model; and they are found to be
mutually consistent.  I have shown that the interaction giving rise
to the fractional exclusion statistics for the elementary excitations
of a given condensed matter system can in fact be treated as the
statistical gauge field carried by the real particles making up
the system.  This was done by calculating the
exact dynamical density-density correlation function
and one-particle Green's function using Jack
symmetric polynomials and examining the intermediate states contributing
to the correlation functions.  I find that the intermediate states
for $\lambda = p/q$ CSM are spanned by $q$ quasiparticles and $p$
quasiholes for the density-density correlation function and
$q-1$ quasiparticles and $p$ quasiholes for the hole propagator
and thereby show that the quasiparticles indeed carry charge 1 and
flux $2\pi\lambda$ and the quasiholes charge $-1/\lambda$ and
flux $-2\pi$ as first suggested by Haldane \cite{duncan94}.

I also construct an explicit multi-sector anyon operators in analogy
with Haldane's harmonic-fluid \cite{duncan82} and calculate
their correlation functions which agree with those of the
CSM which corresponds to the first-quantized construction of anyons.
Therefore, the CSM at odd-integer coupling constant describes
the edge states of the fractional quantum Hall droplet corresponding
to the Laughlin states as suggested in Ref. \cite{csmedge}.

There are some interesting open problems:
\begin{itemize}
\item How can one rigorously construct 1D fractional exchange
statistics using the Yang-Baxter equation ?
\item For the CSM with the coupling constant other than $\lambda = 1/2$
, $1$, and $2$, are there any corresponding random matrices?
\item How can one generalize Jack polynomials to the $SU(n)$ case?
\end{itemize}
I hope to see some of these questions answered in the future.

\section{Acknowledgment}

This work is supported by DOE grant DE-FG02-90ER40542.

\appendix
\section{How to take the thermodynamic limit}

The thermodynamic limit of the DDDCF given by
Eq. (\ref{finited}) greatly simplifies by the fact that
the coefficient $[0']_{\mbox{\boldmath $\kappa$}}^\lambda$ vanishes unless
the diagram $\cal D ({\mbox{\boldmath $\kappa$}})$ has at most
$p$ rows of length longer than $q$ and $q$ columns of length longer than
$p$.  As shown in Section \ref{sec:fractional}
the minimal charge neutral excitations in the CSM consist of
$q$ quasiparticles and $p$ quasiholes;
therefore, the intermediate states that contribute to
the correlation function have
exactly $q$ quasiparticle and $p$ quasihole excitations.\footnote{
The non-vanishing diagrams with less than $p$ rows and
$q$ columns are interpreted as having some of the quasiparticles
and quasiholes in their unexcited modes.}

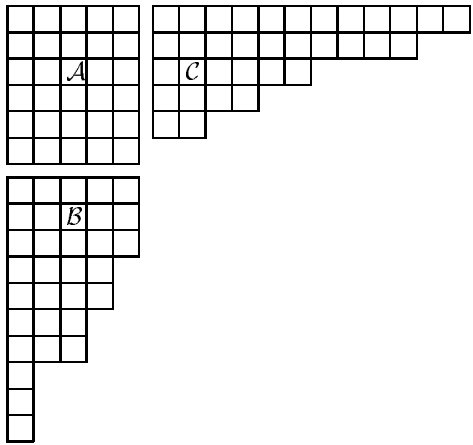
\begin{figure}
\caption{A diagram ${\cal D}$ that contributes to the DDDCF at
$\lambda = p/q = 5/6$ is divided into three subdiagrams
$\cal A$, $\cal B$, and $\cal C$ for convienience.}
\label{subdiagram}
\end{figure}

The contributing diagrams ${\cal D}({\mbox{\boldmath $\kappa$}})$, for
convenience, are divided into three subdiagrams
${\cal A}({\mbox{\boldmath $\kappa$}}) = \{(i,j),\
1\le i \le p,\ 1 \le j \le q\}$,
${\cal B}({\mbox{\boldmath $\kappa$}}) = \{(i,j),\
1\le j \le q,\ q+1 \le i \le \kappa_j'\}$ and
${\cal C}({\mbox{\boldmath $\kappa$}}) = \{(i,j),\
1\le i \le p,\ p+1 \le j \le \kappa_i\}$ as illustrated in
Fig. \ref{subdiagram}.
The factors expressed in terms of the generalized factorial
over ${\cal D}({\mbox{\boldmath $\kappa$}})$ are evaluated for
each of the subdiagrams, separately.  Then, later the each subfactors
are multiplied to obtain to full factors over the diagram $\cal D$.

First, consider the following factor that appears in the DDDCF
\begin{equation}
F_1 = \prod_{(i,j)\in {\mbox{\boldmath $\kappa$}}} \left(
{N + (j-1)/\lambda -(i-1) \over N + j/\lambda - i}\right).
\end{equation}
$F_1$ over ${\cal A}$ simplifies in the thermodynamic limit to
\begin{equation}
F_1({\cal A}) = \prod_{j=1}^p \prod_{i=1}^q \left(
{N + (j-1)/\lambda -(i-1) \over N + j/\lambda - i}\right)
\stackrel{N\rightarrow\infty}{\rightarrow} 1.
\end{equation}
In order to evaluate $F_1$ over ${\cal B}$ I rewrite the product in terms
of the gamma functions using the identity
$\Gamma(z+n)/\Gamma(z) = z(z+1)\cdots(z+n-1)$ as
\begin{eqnarray}
F_1({\cal B}) & = & \prod_{j=1}^p {(N+(j-1)/\lambda-\kappa_j'+1)\cdots
(N+(j-1)/\lambda-q) \over (N+j/\lambda-\kappa_j')\cdots
(N+j/\lambda-q-1)} \nonumber \\
\mbox{} & = & \prod_{j=1}^p {\Gamma(N+j/\lambda-\kappa_j')
\Gamma(N+j/\lambda-q+1-1/\lambda)\over
\Gamma(N+j/\lambda-\kappa'_j+1-1/\lambda)\Gamma(N+j/\lambda-q)}.
\label{f1b}
\end{eqnarray}
Using the following asymptotic relation,
\begin{equation}
\lim_{|z| \rightarrow \infty} {\Gamma(z+a)\over \Gamma(z)} = z^a,
\label{gamma}
\end{equation}
I reduce $F_1({\cal B})$ in Eq.\ (\ref{f1b}) as
\begin{eqnarray}
F_1({\cal B}) &\stackrel{N\rightarrow\infty}{\rightarrow}&
\prod_{j=1}^p \left({N+j/\lambda -q \over N+j/\lambda -\kappa_j'}
\right)^{1-1/\lambda} \nonumber \\
\mbox{} &\rightarrow&
\prod_{j=1}^p (1-\kappa_j'/N)^{1/\lambda-1}
\end{eqnarray}
$F_1$ over ${\cal C}$, similarly, is given by
\begin{equation}
F_1({\cal C}) \stackrel{N\rightarrow\infty}{\rightarrow}
\prod_{i=1}^q \left( 1+ {1\over \lambda}
{\kappa_i \over N}\right)^{\lambda-1}.
\end{equation}

Evaluation of $([0']_{\mbox{\boldmath $\kappa$}}^\lambda)^2$
is straightforward and is equal to product of the following three factors
\begin{eqnarray}
([0']_{\mbox{\boldmath $\kappa$}}^\lambda)^2 ({\cal A}) & = &
\lambda^{-2(p-1)} \Gamma^2(p) \prod_{j=1}^p
{\Gamma^2(q-(j-1)/\lambda)\over \Gamma^2(1-(j-1)/\lambda)}, \\
([0']_{\mbox{\boldmath $\kappa$}}^\lambda)^2 ({\cal B}) & = &
\prod_{j=1}^p {\Gamma^2(-(j-1)/\lambda + \kappa_j') \over
\Gamma^2(-(j-1)/\lambda + q)}, \\
([0']_{\mbox{\boldmath $\kappa$}}^\lambda)^2 ({\cal C}) & = &
\lambda^{-2(\sum_{i=1}^q \kappa_i) +2pq} \prod_{i=1}^q
{\Gamma^2(\kappa_i -\lambda(i-1))\over
\Gamma^2(p-\lambda(i-1))}.
\end{eqnarray}
The value of
$j_{\mbox{\boldmath $\kappa$}}^\lambda$ over ${\cal A}$ is given by
\begin{equation}
j_{\mbox{\boldmath $\kappa$}}^\lambda ({\cal A}) =
\prod_{i=1}^q\prod_{j=1}^p (\kappa_j' + {1\over \lambda} \kappa_i)^2.
\end{equation}

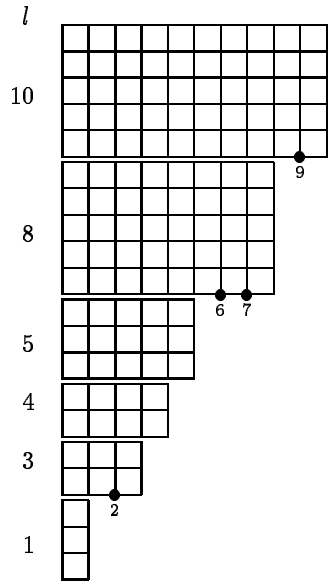
\begin{figure}
\caption{A subdiagram $\cal B$ with $p = 10$ is divided into $p$ {\it cells}.
The dots $\bullet$ indicate the empty cells where
$\kappa'_l = \kappa'_{l+1}$, and only the
$1$, 3, 4, 5, 8, and 10{\it th} cells are non-empty.
For the $l$th cell, $\kappa_i$ are all same and equal to $l$.}
\label{cells}
\end{figure}

In order to evaluate $j_{\mbox{\boldmath $\kappa$}}^\lambda$ over
${\cal B}$, I further divide up the subdiagram ${\cal B}$ into
$p$ cells such that $l$th cell is given by
$\{(i,j),\ 1\le j \le l,\ \kappa_{l+1}'+1 \le i \le \kappa_l' \}$.
If $\kappa_l' = \kappa_{l+1}'$, then the $l$th cell is empty.  Fig.
\ref{cells} illustrates how the subdiagram $\cal B$ is divided into
$p$ cells and how the empty cells appear.
The value of $j_{\mbox{\boldmath $\kappa$}}^\lambda$ over $\cal B$
is then as follow
\begin{eqnarray}
j_{\mbox{\boldmath $\kappa$}}^\lambda ({\cal B}) & = &
\prod_{l=1}^p \prod_{j=1}^l \prod_{i=\kappa_{l+1}'+1}^{\kappa_l'}
(\kappa_j'-i+1+(l-j)/\lambda)(\kappa_j'-i+(l-j+1)/\lambda) \nonumber \\
\mbox{}&=& \prod_{l=1}^p\prod_{j=1}^l
{\Gamma(\kappa_j'-\kappa_{l+1}'+1+(l-j)/\lambda)
\Gamma(\kappa_j'-\kappa_{l+1}' + (l-j+1)/\lambda) \over
\Gamma(\kappa_j'-\kappa_l'+1+(l-j)/\lambda)
\Gamma(\kappa_j'-\kappa_l'+(l-j+1)/\lambda)},
\end{eqnarray}
where $\kappa_{p+1}' \equiv q$.  The contributions from the
empty cells are identically equal to one. For the non-empty
$l$th cell $\kappa_i = l$ for $\kappa_{l+1}'+1 \le i \le \kappa_l'$.
The expression above simplifies further to
\begin{eqnarray}
j_{\mbox{\boldmath $\kappa$}}^\lambda ({\cal B}) & = &
{1\over \Gamma^p(1/\lambda)}
\prod_{j=1}^p \Gamma(\kappa_j'+1-j/\lambda)
\Gamma(\kappa_j'-(j-1)/\lambda) \nonumber \\
\mbox{} &\times& \prod_{i>j}
{\Gamma(\kappa_i'-\kappa_j'+1-(i-j+1)/\lambda)
\Gamma(\kappa_i'-\kappa_j'-(i-j)/\lambda) \over
\Gamma(\kappa_i'-\kappa_j'+1-(i-j)/\lambda)
\Gamma(\kappa_i'-\kappa_j'-(i-j-1)/\lambda)}.
\end{eqnarray}
In order to take the $N\rightarrow \infty$ limit using Eq.
(\ref{gamma}), I take the following
ratio first and then send the ratio to the limit,
\begin{equation}
{([0']_{\mbox{\boldmath $\kappa$}}^\lambda)^2 ({\cal B}) \over
j_{\mbox{\boldmath $\kappa$}}^\lambda ({\cal B})}
\stackrel{N\rightarrow\infty}{\rightarrow}
{\Gamma^p(1/\lambda)\over \prod_{j=1}^p \Gamma^2(q-(j-1)/\lambda)}
\prod_{j=1}^p (\kappa_j')^{1/\lambda-1}
\prod_{i>j} |\kappa_i'-\kappa_j'|^{2/\lambda}.
\label{jb}
\end{equation}
Similarly, $j_{\mbox{\boldmath $\kappa$}}^\lambda$ over ${\cal C}$ is
given by
\begin{eqnarray}
j_{\mbox{\boldmath $\kappa$}}^\lambda ({\cal C}) & = &
{\lambda^{-2(\sum_{i=1}^q\kappa_i)+2pq} \over
\Gamma^q(\lambda)}
\prod_{i=1}^q \Gamma(\kappa_i-\lambda(i-1))
\Gamma(\kappa_i+1-\lambda i) \nonumber \\
\mbox{} &\times & \prod_{j>i} {
\Gamma(\kappa_i-\kappa_j - \lambda(i-j))
\Gamma(\kappa_i-\kappa_j+1-\lambda(i-j+1)) \over
\Gamma(\kappa_i-\kappa_j-\lambda(i-j-1))
\Gamma(\kappa_i-\kappa_j+1-\lambda(i-j))}.
\end{eqnarray}
I take the following ratio and sent it to the thermodynamic limit using
Eq. (\ref{gamma}),
\begin{equation}
{([0']_{\mbox{\boldmath $\kappa$}}^\lambda)^2 ({\cal C}) \over
j_{\mbox{\boldmath $\kappa$}}^\lambda ({\cal C})}
\stackrel{N\rightarrow\infty}{\rightarrow}
{\Gamma^q(\lambda)\over \prod_{i=1}^q \Gamma^2(p-\lambda(i-1))}
\prod_{j=1}^p \kappa_j^{\lambda-1}
\prod_{i>j} |\kappa_i-\kappa_j|^{2\lambda}.
\label{jc}
\end{equation}
Putting all the terms together, setting $\kappa_j/N = x_j$ and
$\kappa_j'/N = y_j$ and turning the sums into integrals
I get Eq. (\ref{ddc}) with the normalization constant given by
Eq. (\ref{norm}).

If the differences $|\kappa'_i - \kappa'_j|$
($|\kappa_i - \kappa_j|$) are of order ${\cal O}(1)$ then the corresponding
contributions to Eq. (\ref{jb}) (Eq. (\ref{jc}))  are of order
${\cal O}(1)$ instead of ${\cal O}(N^{2/\lambda})$ (${\cal O}(N^{2\lambda})$)
as $N\rightarrow \infty$; therefore, the contributions
of such terms to the DDDCF are suppressed in the thermodynamic limit.
If $\kappa_j'$ is of order ${\cal O}(1)$ or ${\cal O}(N)$ then there
is a corresponding contribution of order ${\cal O} (N^{1-1/\lambda})$;
so, if $\lambda < 1$ the form factor vanishes as $N\rightarrow \infty$,
and if $\lambda > 1$ it diverges.
However, the net contribution of such term
to the DDDCF is of order $N^{-1} \times N^{1-1/\lambda}$ which
always vanishes so long as $\lambda \ge 0^+$. (The factor $1/N$ comes from
$\Delta y_j$.) More generally,
let there be $r$ $\kappa_j'$ of all order ${\cal O}(1)$ or ${\cal O}(N)$.
Then, the form factor will be of order $N$ to the power of
$r(1-1/\lambda)-(2/\lambda)r(r-1)/2 = r(1-r/\lambda)$.  Therefore,
if the $\lambda>r$ the form factor diverges. In other words such
configuration is favorable.  This is an exotic violation of the
super selection rule since $r$ quasiholes are stuck together.
However, the net contributions of these terms are again vanishingly
small except, of course, in the long-wavelength limit where only these
exceptional states survive.

Almost identical method can be applied to obtain Eq. (\ref{propagator}).
When the limit $x, t \rightarrow 0$ is taken, the propagator becomes
just the static density $\rho_0$. This gives a generalization of the
Selberg's integral formula and might have connections with the
$q$-deformed Lie algebra.

\section{Asymptotic Expansion}

\subsection{Dynamical density-density correlation function}

A method for expanding the dynamical density-density
correlation functions as $x \rightarrow \infty$
is presented in the appendix. A similar method has been used previously
in \cite{forrester} for the equal-time correlation functions.

The form factor to the DDDCF is largest near a phase space region where
all $x_j$'s near zero and $y_i$'s near zero or one.  Thus, the integrand
will be expanded about this region.  In Fig. \ref{sector}
the shaded region gives
non-zero contributions to the DDDCF and the regions with darker shades
give largest contributions to the DDDCF in the long-wavelength limit.
Each of these dark regions is  labeled with the index $n$ and is called
$n$th sector, whose contribution to the correlation
function has a characteristic oscillation, and is spanned by
all $x_i$'s close to zero and $n$ ($p-n$) $y_j$'s close to one (zero).

To leading order in the normalized momentum variables
the following relation hold true
\begin{equation}
Qx \pm Et = 2\pi\rho_0 \xi^\mp \left(\sum_{i=1}^q x_i +\sum_{j=1}^{p-n}
y_j\right) - 2\pi\rho_0 \xi^\pm \sum_{k=1}^n w_k + 2\pi\rho_0 n x,
\end{equation}
where $\xi^\pm = x \mp v_s t$.  The sound velocity $v_s$ is
$2\pi\rho_0 \lambda$.  A new set of $n$ variables
$w_k$ for $y_j$ close to one are introduced such that $w_k = 1 - y_j$.
$\xi^+$ ($\xi^-$) is the space-time
coordinate for the right (left) moving particle or hole excitations.

To leading order in each of the $p$ sectors the DDDCF is given by
\begin{eqnarray}
\langle\rho(x,t)\rho(0,0)\rangle &\approx& \frac{C}{2}
\left(I_1^\lambda(+i|q,p)\left({1\over \xi^-}\right)^2  +
I_1^\lambda(-i|q,p)\left({1\over \xi^+}\right)^2 \right)
\nonumber \\
\mbox{} & + & \sum_{n = 1}^p \tilde{\cal C}_n
\left({1\over \xi^+\xi^-}\right)^{n^2\over \lambda} \frac{1}{2}
\left(I_1^\lambda(-i|q, p-n) I_2^\lambda(+i|n) e^{i2\pi\rho_0 n x}
\right. \nonumber \\
\mbox{} & + &
\left. I_1^\lambda(i|q, p-n) I_2^\lambda(-i|n) e^{-i2\pi\rho_0 n x}\right),
\label{add}
\end{eqnarray}
where $\tilde{\cal C}_n = {p \choose n} C
\lambda^{-2nq} n^2 \left({1\over 2\pi\rho_0}\right)^{{2n^2\over \lambda}-2}$
and the functions $I_1^\lambda$ and $I_2^\lambda$ are defined as follow
\begin{eqnarray}
I_1^\lambda (z|l,p-m) &=& \prod_{i=1}^l \int_0^\infty dx_i
\prod_{j=1}^{p-m} \int_0^\infty dy_j
\left\{ \begin{array}{ll}
          \left(\sum_{i=1}^l x_i +
          \sum_{j=1}^p y_j \right)^2  & \mbox{if $m = 0$} \\
          1  & \mbox{if $m \ne 0$}
        \end{array}
\right\} \nonumber \\
\mbox{} &\times & \prod_{i=1}^l \prod_{j=1}^{p-m} (x_i + \lambda y_j)^{-2}
{\prod_{i<j} |x_i-x_j|^{2\lambda}
\prod_{i'<j'} |y_{i'}-y_{j'}|^{2/\lambda} \over
\prod_{i=1}^l x_i^{1-\lambda}
\prod_{j=1}^{p-m} y_j^{1-1/\lambda}}  \nonumber \\
\mbox{} & \times & \exp\left(-z\left(\sum_i x_i +
\sum_j y_j\right)\right)  \\
I_2^\lambda (z|m) & = & \prod_{k=1}^m \int_0^\infty dw_k
{\prod_{i<j} |w_i - w_j|^{2/\lambda} \over
\prod_{k=1}^m w_k^{1-1/\lambda}}
\exp\left(-z\sum_k w_k\right)
\end{eqnarray}

Since $I_1^\lambda$ and $I_2^\lambda$ are absolutely
convergent only if $Re(z) > 0$, it
is necessary to analytically continue the functions as follow
\begin{eqnarray}
I_1^\lambda (z|l,m) & = &
\left({1\over z}\right)^{\lambda l^2 + m^2/\lambda -2lm + 2\delta_{m,p}}
I_1^\lambda (1|l,m), \\
I_2^\lambda (z|m) & = & \left({1\over z}\right)^{m^2/\lambda}
I_2^\lambda (1|m).
\end{eqnarray}
The above analytical extensions give the following relations:
$I_1^\lambda(-i|q,p-n) I_2^\lambda(+i|n) =
I_1^\lambda(1|q,p-n) I_2^\lambda(1|n)$ and $I^\lambda_1(\pm i|q,p)
= - I^\lambda_1(1|q,p)$.

Finally, the DDDCF to leading order in each harmonic mode is given by
\begin{equation}
\langle 0|\rho(x,t)\rho(0,0)|0\rangle \approx
- \frac{C}{2} I_1^\lambda(1|q,p)
\left({1\over (\xi^+)^2} + {1\over (\xi^-)^2} \right)
+ \sum_{n=1}^p {\cal C}_n
\left({1\over \xi^+ \xi^-}\right)^{n^2/\lambda} \cos(2\pi \rho_0 n x),
\label{finaladd}
\end{equation}
where ${\cal C}_n = \tilde {\cal C}_n I_1^\lambda(1|q,p-n)
I_2^\lambda(1|n)$.

The renormalization parameter is defined in
\cite{duncan82} as $\eta = 2(v_J/v_N)^{1/2}$, where $v_J$ and
$v_N$ are the current and charge velocities. For the $U(1)$ CSM
$\eta = 2/\lambda$ \cite{hahal92}.\footnote{The interaction
coupling constant $\lambda$ used in Ref. \cite{hahal92} corresponds
to $\lambda - 1$ in this paper.}
With this identification the form of Eq. (\ref{finaladd}) in the
static limit agrees with the expansion given in
\cite{duncan82}.  Furthermore, by matching the universal
constant in the $0$th sector I deduce the following
integral formula
\begin{equation}
I_1^\lambda(1|q,p) = (2\pi^2 \lambda C)^{-1}.
\end{equation}

\subsection{One-particle Green's function}

Using the similar method used for the DDDCF I find the
following leading order terms for the hole propagator
\begin{equation}
\langle 0|\Psi^\dagger(x,t)\Psi(0,0)|0\rangle \approx
\sum_{n=0}^p {\cal D}_n
\left({1\over 2\pi \rho_0 \xi^+}\right)^{(n-\lambda)^2/\lambda}
\left({1\over 2\pi \rho_0 \xi^-}\right)^{n^2/\lambda}
e^{i(2\pi\rho_0(n-\lambda/2)x + (\pi\rho_0\lambda)^2 t)},
\end{equation}
where ${\cal D}_n = \rho_0 D \lambda^{1-\lambda-2(q-1)n}
{p\choose n} I_1^\lambda(1|q-1,p-n) I_2^\lambda(1|n)
e^{-i\pi(n-\lambda/2)}$.  Here, $I_1^\lambda(z|q-1,p)$ is
defined as before but without the pre-factor $\left(
\sum_i x_i + \sum_j y_j \right)^2$.

\begin{figure}
\caption{The shaded area is the region of support for the dynamical
density-density correlation function at $\lambda = p/q = 5/3$
in the energy-momentum space. The energy $E$ is in some
arbitrary unit and the momentum $Q$ in units of $2\pi \rho_0$.
The $2p$ distinct regions with darker shade corresponds to
the low-energy sectors of the model.}
\label{sector}
\end{figure}

\begin{figure}
\caption{Cylinderical geometry for the anyon system.  Two oppositely-moving
edges along $x$-axis are separated by an incompressible anyon fluid.
The cylinder is pierced by a solenoid which induces a magnectic flux
$\Phi$ through its cross-section. }
\label{cylinder}
\end{figure}

\vfill\eject
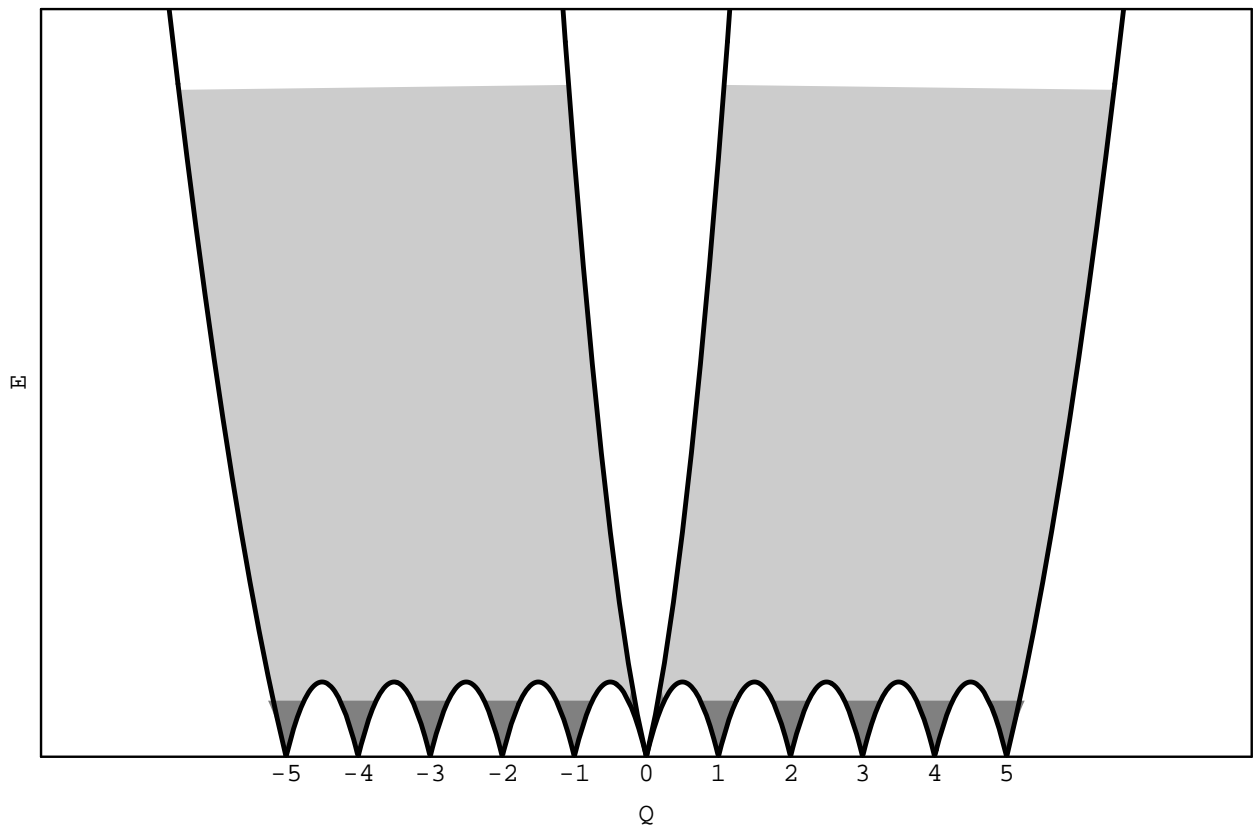
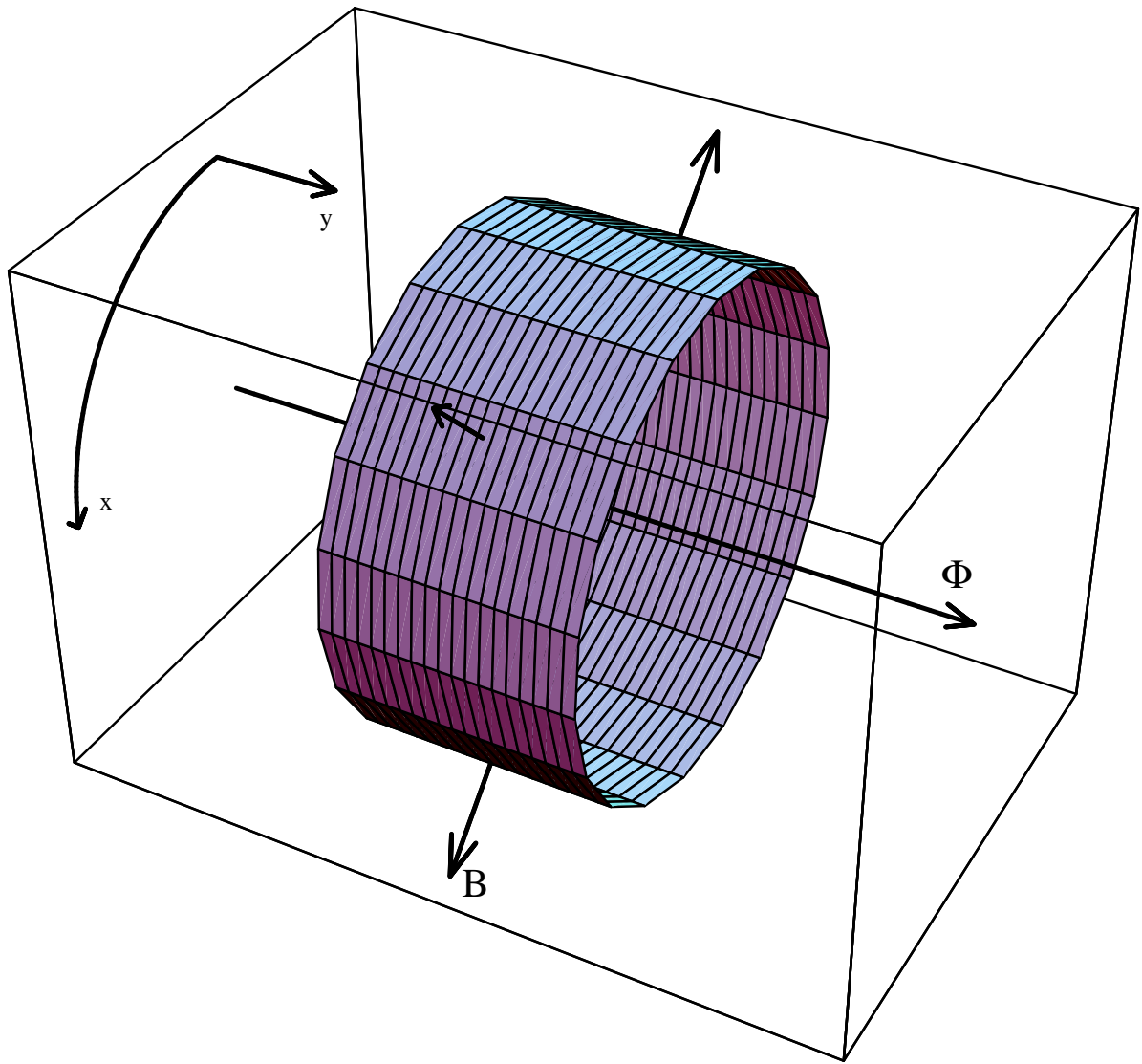
\end{document}